%% file: depth.tex
\documentclass[12pt,english]{article}

\newcommand{\blind}{1}

\include{preamble}

\myexternaldocument{depthSupp}%

\def\references{\bibliography{zotero.bib}}

\begin{document}

\def\spacingset#1{\renewcommand{\baselinestretch}%
{#1}\small\normalsize} \spacingset{1}

%%%%%%%%%%%%%%%%%%%%%%%%%%%%%%%%%%%%%%%%%%%%%%%%%%%%%%%%%%%%%%%%%%%%%%%%%%%%%%

\if1\blind
{
  \title{\bf Tukey's Depth for Object Data}
  \author{Xiongtao Dai\thanks{
    Partially supported by NSF grant DMS-2113713. Email: \texttt{xdai@iastate.edu} }\hspace{.2cm}\\
    Department of Statistics, Iowa State University, Ames, Iowa 50011 USA \\
    and\\
    Sara Lopez-Pintado\thanks{
    Partially supported by NSF grant DMS-2113713. The author would also like to thank funding provided by NIH grant 1R21 MH120534-01. Email:  \texttt{s.lopez-pintado@northeastern.edu}}\hspace{.2cm}\\
    Department of Health Sciences, Northeastern University, Boston, MA 02115 USA\\\\
    for the Alzheimer’s Disease Neuroimaging Initiative\thanks{Data used in preparation of this article were obtained from the Alzheimer’s Disease Neuroimaging Initiative (ADNI) database (adni.loni.usc.edu). As such, the investigators within the ADNI contributed to the design and implementation of ADNI and/or provided data but did not participate in analysis or writing of this report. A complete listing of ADNI investigators can be found at: \url{http://adni.loni.usc.edu/wp-content/uploads/how_to_apply/ADNI_Acknowledgement_List.pdf}}
    }
  \maketitle
} \fi

\if0\blind
{
  \bigskip
  \bigskip
  \bigskip
  \begin{center}
    {\LARGE\bf Tukey's Depth for Object Data}
\end{center}
  \medskip
} \fi

\bigskip
\begin{abstract}
We develop a novel exploratory tool for non-Euclidean object data based on data depth, extending the celebrated Tukey's depth for Euclidean data. The proposed metric halfspace depth, applicable to data objects in a general metric space, assigns to data points depth values that characterize the centrality of these points with respect to the distribution and provides an interpretable center-outward ranking.
Desirable theoretical properties that generalize standard depth properties postulated for Euclidean data are established for the metric halfspace depth. The depth median, defined as the deepest point, is shown to have high robustness as a location descriptor both in theory and in simulation.
We propose an efficient algorithm to approximate the metric halfspace depth and illustrate its ability to adapt to the intrinsic data geometry.
The metric halfspace depth was applied to an Alzheimer's disease study, revealing group differences in the brain connectivity, modeled as covariance matrices, for subjects in different stages of dementia.
Based on phylogenetic trees of 7 pathogenic parasites, our proposed metric halfspace depth was also used to construct a meaningful consensus estimate of the evolutionary history and to identify potential outlier trees.
\end{abstract}

\noindent%
\vfill

\newpage
\spacingset{1.45} % DON'T change the spacing!

\section{Introduction\label{sec:introduction}}

\subsection{Backgrounds\label{ssec:backgrounds}}
Complex data objects are increasingly generated across science and rapidly gaining relevance. 
Finite-dimensional non-Euclidean data is an important class of object data \citep{marr:14}, which models, for example, directions \citep{mard:09}, covariance matrices \citep{penn:06}, and trees \citep{bill:01}.  There has been extensive development in methods and theory to address the complexity of these objects, including location measures  \citep{frec:48}, statistical inference \citep{bhat:05}, and classification \citep{dai:18-5}.
However, exploratory data analysis is a crucial paradigm that lacks development for these nonstandard data.
As the basic data units become more complex and multifaceted, there is an escalated need for an agnostic exploratory data analysis.
Data exploration before modeling will reveal properties of the data distribution and help identify extreme versus typical observations. In this regard, a first step is to overcome the absence of a canonical ordering for complex objects and propose principled definitions of rank, median, and order statistics. 

Data depth has been proven to be a powerful exploratory and data-driven tool that can be used to rank observations and reveal features of the underlying data distribution. The notion of data depth was originally introduced for multivariate Euclidean data and provides a way of measuring how “representative” or “outlying” an observation is with respect to a probability distribution. In particular, a depth function assigns a non-negative depth value to a given observation within a distribution, where the larger this value is the more central/deep the observation is within the distribution. Points with low depth values correspond to observations near the outskirt of the distribution and ``far'' from the center. These observations could be potential outliers worthwhile of investigation.  Hence, a notion of depth provides a ``center-outward'' ordering for a sample of multivariate observations and allows generalization of ranks, order statistics, central regions \citep[see][]{zuo:00-1}, and robust inferential and classification methods to multivariate data \citep{li:12}. 

For multivariate Euclidean data, Tukey's halfspace depth \citep{tuke:75} has attracted much attention. Due to its intuitive properties \citep{zuo:00-1} and robustness of the depth induced median \citep{dono:92}, Tukey's depth stands out as the first and most popular depth among a rich body of depth notions proposed for multivariate data \citep[e.g.,][]{oja:83,liu:90,einm:15}. It not only leads to an intuitive center-outward ranking for multivariate data, but also enables the development of graphical data summaries \citep{tuke:75,rous:99} and robust nonparametric rank tests \citep{liu:93-1,chen:12-2}. The classical Tukey's depth, however, relies on the Euclidean geometry and is inappropriate for non-Euclidean data objects. %a naive application of the Tukey's depth to data lying on a sphere will result in zero depth for all points. 

%However, Tukey's depth cannot be easily extended to non-Euclidean.
%There are hardly any depth notions for these objects that satisfy as many desirable properties as Tukey's depth, so further development is much needed. 

%One of the first and most popular depths for multivariate data is Tukey’s depth or also known as halfspace depth  (Tukey, 1975). It %satisfies the desirable depth properties established in Zuo and Serfling 2000. 

% Tukey depth is also known as location depth or halfspace depth. Given a finite set S of n points and a point p in , the Tukey depth of p is defined as the minimum number of points of S contained in any closed halfspace with p on its boundary [15], [22]. An equivalent definition is the minimum number of points of S contained in any halfspace which also contains p

{
Though defining depth notions for non-Euclidean data has garnered wide interest,
the literature has focused on specialized spaces, such as a unit sphere
\citep{smal:87,liu:92,pand:18}, positive definite matrices
\citep{flet:11-2,chau:19}, networks \citep{frai:17}, data on a graph
\citep{smal:97}, and infinite-dimensional functional data \citep{frai:01,lope:09}.
\citet{chen:18-3} and \citet{pain:18} considered halfspace depth for the scatter matrix
of Euclidean data points. 
\citet{frai:19} proposed a spherical depth that applies to Riemannian manifold data. 
Targeting general settings, \citet{carr:96} sketched a halfspace depth based on dissimilarity measures without methodological development, although this depth is different from ours in general; see \autoref{sec:MHDC96} in the Supplemental Materials. 
\cite{carr:96} also introduced an extension of the halfspace depth to a regression setting closely related to the regression depth proposed by \cite{rous:99-2}; see also \cite{zuo:21} for a discussion of the theoretical properties of regression depths. 
}

\subsection{Our Contributions\label{ssec:contribution}}

The goal of our work is to generalize Tukey's depth to data objects taking values on an arbitrary metric space, defining a general depth notion that shares the desirable properties of Tukey's depth. 
This will make available depth-based exploratory and robust inferential toolsets to general object data. 

%Non-Euclidean data objects are increasingly common as the types of data generated and collected become richer. 
%%An object-oriented perspective \citep{marr:14} for analyzing non-Euclidean data is to view them as members of a non-Euclidean object space and analyze them based on operations defined in the object space. 
%The strength and applicability of our proposed depth and related methods will be illustrated on two complex data sets consisting on: a) connectivity matrices from functional magnetic resonance imaging (fMRI) data of patients with dementia and healthy controls and b) phylogenetic trees comparing the genetic materials from different species. 

%
\begin{comment}
on which the neighborhood of each point resembles bended Euclidean
subsets with a common dimension. 
\end{comment}

We propose the metric halfspace depth in \ref{ssec:MHD}, which is a generalization of Tukey's depth to object data on a general metric space. 
Metric halfspace depth incorporates the data space geometry through the distance metric
$d$, a feature available on any metric space. 
The proposed metric halfspace depth, therefore, applies to a wide range of non-Euclidean data objects. 
This includes data lying on smooth Riemannian manifolds such as directional data on a sphere \citep{mard:09}; bivariate molecular torsion angles on a flat torus \citep{eltz:18}; and constrained matrix-valued data, such as rotations \citep{bing:09} and covariance matrices \citep{dai:20}. 
Nonsmooth objects with possible degeneracy lying on a geodesic space, such as phylogenetic trees \citep{fera:20}, networks \citep{kola:20}, and shapes \citep{dryd:16} can also be investigated by the proposed depth.
%A commonality in all these data objects is that in each case, there exists a distance metric between data objects that induces the geometry and gives rise to proximity, neighborhood, and halfspaces. 
%The vast majority of data analysis are considered in metric spaces with additional structures, such as the existence of a geodesic, i.e., a shortest path between each pair of points. 

%The metric halfspace depth does not rely on linearization to tangent spaces and therefore defined for any data distribution lying on non-smooth spaces beyond Riemannian manifolds, whereas linearization techniques are limited to concentrated data lying on smooth manifolds.
%No moment condition is required for the metric halfspace depth, in analogy to the classical Tukey's depth. 

We establish desirable properties of the metric halfspace depth in
\ref{sec:properties}, extending much of the properties enjoyed by the classical Tukey's
depth \citep{tuke:75} for Euclidean data. The axiomatic properties
of depth notions introduced in \citet{zuo:00-1} are satisfied to a great
extent in many commonly investigated data spaces. The metric halfspace
depth is invariant to a large class of transformations; if the data
are symmetrically distributed around a center, then the center has
the maximal metric halfspace depth; the depth values have a center-outward tendency and monotonically decrease from the deepest point to the peripheral points; 
and the depth vanishes as one moves
away from the center. The metric halfspace depth function is upper
semi-continuous, which implies that the nested deepest regions are compact. 
We establish a root-$n$ rate of convergence of the sample depth to the true depth function, and the consistency of the sample deepest point to the population deepest point, {assuming uniqueness of the latter}.
Moreover, the metric halfspace depth is shown to be robust to contamination, having a high breakdown point for symmetric distributions regardless of the dimension of the data space. 
All proofs are included in the Supplemental Materials.

The classical Tukey's depth has a well-known weakness in its high computation cost even in moderate dimensions. 
To overcome this obstacle, we propose efficient algorithms in \ref{sec:computation} to approximate the metric halfspace depth by looking into finitely many halfspaces as informed by the
dataset. 
%The proposed algorithm, in the special case of a Euclidean space, adds to the approximation algorithms \citep{rous:98,cues:08,bogi:18} for calculating the Tukey's depth.
Our proposed {approximation} algorithm for calculating the depth function and the deepest point has a complexity of $O(n^{3})$ with respect to the sample size $n$, independent of the dimension of the data space. 
The approximation algorithm is able to achieve arbitrary precision to the truth by densening the discretization of the space, which we establish in our theoretical results and demonstrate in simulation studies. 
The proposed depth is shown to have excellent numerical performance
in terms of efficiency and robustness in \ref{sec:numericalExperiments}.
The approximate depth algorithm respects the intrinsic data geometry independent of the ambient space as demonstrated in \autoref{asec:numericalApproximationProperties}.
%, a desirable numerical property that does not hold for other approximation algorithms. 

We showcase the practical relevance of the metric halfspace depth in two applications in \ref{sec:applications}, which include (a) neuro-connectivity matrices from functional magnetic resonance imaging (fMRI) data of patients with dementia and healthy controls and (b) phylogenetic trees comparing the genetic materials from different species. 
The application to fMRI data discovered differences in the brain connectivity among groups of normal controls and patients at different dementia stages progressing to Alzheimer's disease using depth-based rank tests. 
The second application considers estimating the phylogenetic history of seven Apicomplexan species, which are pathogenic parasites, in the tree space of \citet{bill:01}. 
We obtained the most representative tree for estimating a consensus evolutionary history of the Apicomplexa and also identified outliers in the individual gene trees. 

\begin{comment}
\begin{itemize}
\item \citet{pain:18} considered halfspace depth for the scatter matrix
of Euclidean data points $\cX$. Although the scatter matrix of a
distribution lie on the manifold SPD (of symmetric positive definite
matrices), their setting appears more like estimating the scatter
out of Euclidean data points. In our setup, we consider a collection
of scatter matrices, e.g. the brain connectivity covariance matrices
for multiple subjects. 
\item \citet{chau:19} proposed manifold zonoid and geodesic depths. The
paper considers the specific manifold of SPD (which they call Hermitian
Positive Definite Matrices), although under some regularity conditions
I think their method can be generalized onto a good-behaving Riemannian
manifold. A crucial weakness of the zonoid depth is that it depends
on the logarithm map to be well-defined. This is the case only for
SPD, which has a cone-structure, but not in general. For example,
on a sphere $S^{d}$, the logarithm map $\log_{p}(-p)$ of antipodals
$(p,-p)$ is undefined. Their geodesic depth is always defined on
any Riemannian manifold, however.
\item \citet{flet:11-1} defines halfspaces for symmetric positive definite
matrices based on Busemann functions, but this construction only works
for non-positively curved Riemannian manifolds.
\end{itemize}
\end{comment}

\section{Metric Halfspace Depth}
\subsection{General Definition \label{ssec:MHD}}

We consider extending the concept of data depth to data objects taking values on a general metric space. 
{
Let $\cM$ be a metric space equipped with distance $d$, and $X$ be an $\cM$-valued random object defined on probability space $(\Omega, \cF, P)$ measurable with respect to the Borel $\sigma$-algebra $\cB(\cM)$. 
}
To define a halfspace depth, the key lies in suitably generalizing the notion of halfspaces. 
For two points $x_{1},x_{2}\in\cM$
on the metric space, we denote the \emph{metric halfspace}, or \emph{halfspace} for brevity, as 
\begin{equation}
\Honetwo=\left\{ y\in\cM\mid d(y,x_{1})\le d(y,x_{2})\right\} ,\label{eq:halfspace}
\end{equation}
which is said to be \emph{anchored at} $(x_{1},x_{2})$. 
Halfspace $\Honetwo$ contains all points of $\cM$ that lie no further
away from $x_{1}$ than from $x_{2}$. Let $\cH=\{\Honetwo\mid x_{1}\ne x_{2}\in \cM\}$
be the collection of all halfspaces and $\cH_{x}=\{\Honetwo\in\cH\mid x\in\Honetwo\}$
the set of halfspaces containing $x$, understanding that the same
halfspace may arise from different pairs of anchors. 
The proposed \emph{metric halfspace depth (MHD)} at $x\in\cM$ w.r.t. {the probability measure $P_X$ induced by $X$} is defined as 
\begin{align}
D(x)=D(x;P_X) & =\inf_{H\in\cH_{x}}{P_X(H)}\label{eq:D1}\\
 & =\infxoxt P(d(X,x_{1})\le d(X,x_{2})).\label{eq:D2}
\end{align}
{
Depth $D(x)$ is the least probability measure of the halfspaces
containing $x$, which is well-defined since the halfspaces are closed and thus measurable. 
}%
Analogously, given i.i.d. observations $X_{1},\dots,X_{n}\in\cM$,
the sample metric halfspace depth at $x\in\cM$ w.r.t. the empirical
distribution $P_{n}$ is 
\begin{align}
D_{n}(x)=D(x;P_{n}) & =\inf_{H\in\cH_{x}}P_{n}(H)\nonumber \\
 & =\infxoxt n^{-1}\sumin\one\{d(X_{i},x_{1})\le d(X_{i},x_{2})\},\label{eq:Dsamp}
\end{align}
where $I\{\cdot\}$ is the indicator function.

It is immediately seen that if $\cM$ is a Euclidean space, then each halfspace
is a closed Euclidean halfspace of the form $\{x\in\bbR^{m}\mid x^{T}v\le c\}$
for some vector $v\in\bbR^{m}$ and $c\in\bbR$, and the metric halfspace
depth coincides with the Tukey's halfspace depth \citep{tuke:75}.
The metric halfspace depth specializes to the angular Tukey's depth
proposed by \citet{liu:92} for data lying on a sphere. 
The metric halfspace depth $D(x)$ captures the geometry of a general metric space $\cM$ through the halfspaces defined by the distance metric $d$. 

The proposed metric halfspace depth measures how central or representative an observation is with respect to the distribution. 
In the context of social choice \citep{capl:88-1},
a point $x\in\cS$ is an ideology, i.e., the favorite proposal shared
by a group of voters. Given two proposals $x_{1}$ and $x_{2}$, ideology $x$ prefers the one closer to itself under distance $d$.
The halfspace probability $P_X(\Honetwo)$ is the proportion
of votes received by proposal $x_{1}$ when posed against $x_{2}$.
Depth value $D(x)$ is 
%the representativeness of $x$ in terms of conformity, which equals 
the least popularity of a proposal that would appeal to $x$; 
in other words, $x$ will not favor an unpopular proposal that wins less than $D(x)$ proportion of votes.
{A related interpretation in facility location problems for a different depth definition can be found in \cite{carr:96}.}

% \section{Halfspace Depth in Metric Spaces\label{sec:depthInMetricSpace}}

\subsection{Metric Spaces\label{ssec:generalNonEuclidean}}

%, such as a \emph{geodesic (metric) space} where there exists a geodesic, i.e., a shortest path between each pair of points; or a \emph{Riemannian manifold}, which is a smooth geodesic space that locally resembles bended Euclidean subsets. 

%In practice, the vast majority of continuous data are analyzed in
%metric spaces with additional interpretable geometric features, such
%as the existence of geodesics between data points which generalizes
%the notion of straight lines. 

%We primarily focus on metric spaces that are complete and connected.
A map $\gamma$ from a closed interval $I\subset\bbR$ to $\cM$ is
said to be a \emph{geodesic} if there exists a constant $\lambda$
such that $d(\gamma(t),\gamma(t'))=\lambda\left|t-t'\right|$ for
all $t,t'\in I$; if further $\lambda=1$, then $\gamma$ is said
to be a \emph{unit speed geodesic}. We say that a geodesic $\gamma$ joins
$x\in\cM$ to $y\in\cM$ if $I=[0,l]$, $\gamma(0)=x$, and $\gamma(l)=y$
for some constant $l$. 
%A \emph{local geodesic} in $\cM$ is a map
%$\gamma$ from an interval $I\subset\bbR$ to $\cM$ such that for
%any $t\in I$, there exists $\epsilon>0$ such that the restriction
%of $\gamma$ to $[t-\epsilon,t+\epsilon]$ is a geodesic. 
%DO WE NEED THE DEFINITION OF LOCAL GEODESIC HERE? 
Now, $(\cM,d)$
is said to be a \emph{geodesic space }if any two points $x,y\in\cM$
are joined by a geodesic. 
\emph{Riemannian manifolds} are smooth submanifolds embedded in an ambient Euclidean space.
{The definitions of the manifolds and additional geometrical quantities, such as the tangent space $T_x\cM$ and exponential map $\exp_x$, are reviewed in \ref{asec:riemannianManifold}. }
The distance between two points $x,y$ on a Riemannian
manifold $\cM$ is the length of the shortest path on $\cM$ connecting
them. 
%For example, a unit sphere, the
%collection of symmetric positive definite matrices, rotation matrices,
%and a flat torus are all complete connected Riemannian manifolds.
Riemannian manifolds are geodesic spaces by the Hopf--Rinow theorem
\citep{lee:18}. 

\begin{figure}[h]
\centering
\resizebox{.7\textwidth}{!}{%
\begin{minipage}[b][1\totalheight][t]{0.7\textwidth}%
\includegraphics[width=1\textwidth]{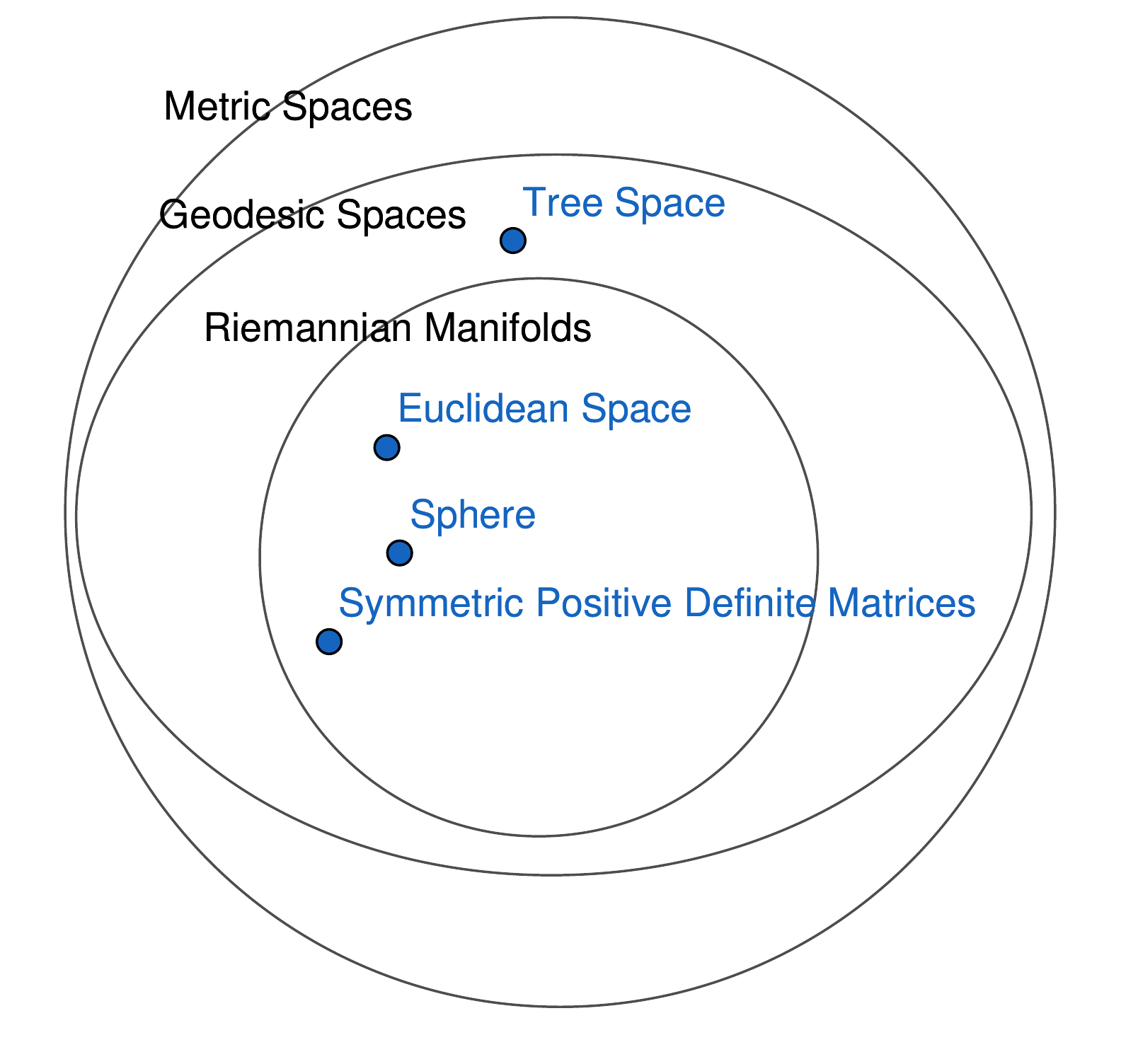}%
\end{minipage}\hfill{}%
\begin{minipage}[b][1\totalheight][t]{0.29\textwidth}%
\noindent\begin{minipage}[t]{1\textwidth}%
\includegraphics[viewport=480bp 0bp 1050bp 680bp,clip,width=1\textwidth]{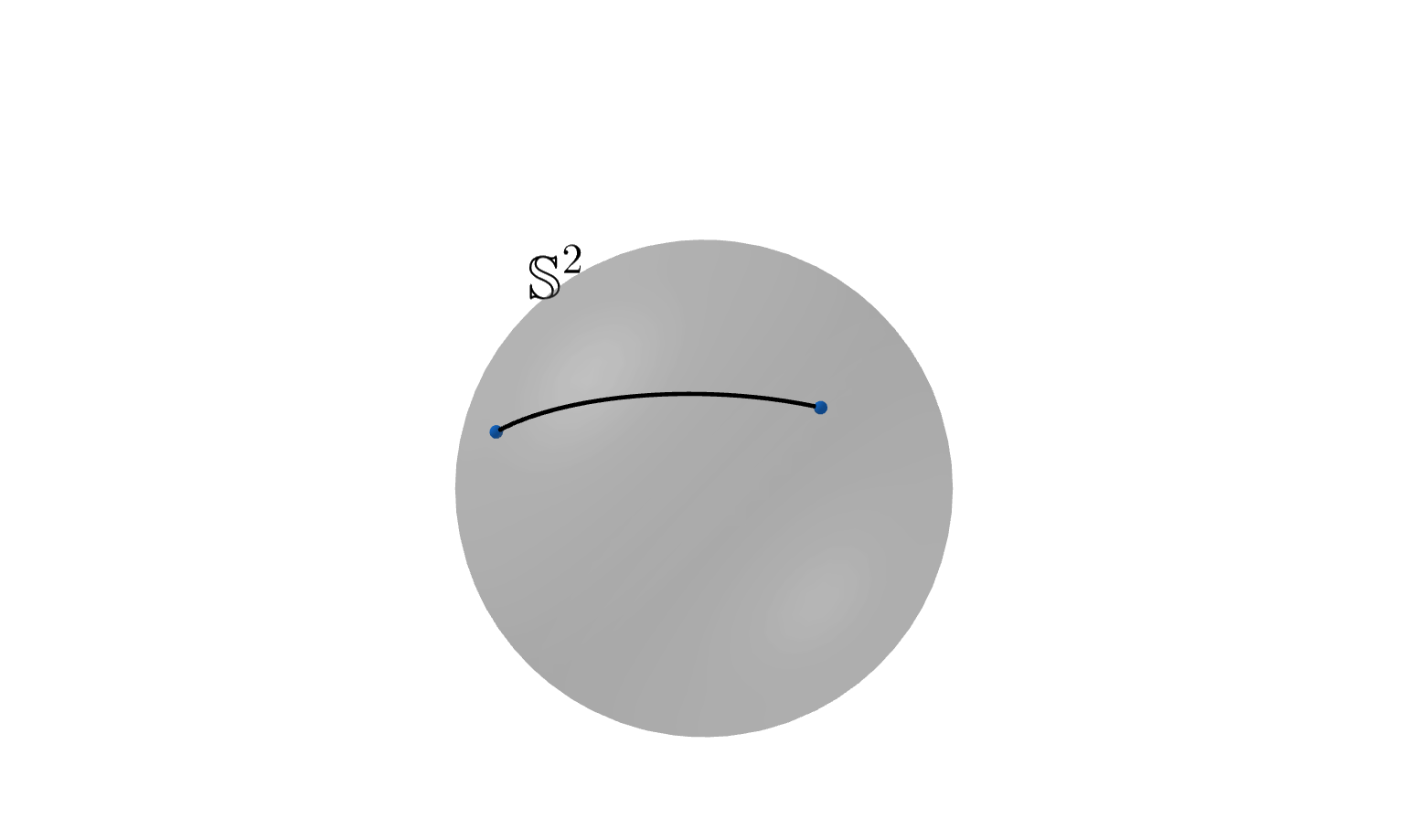}%
\end{minipage}

\noindent\begin{minipage}[t]{1\textwidth}%
\includegraphics[width=1\textwidth]{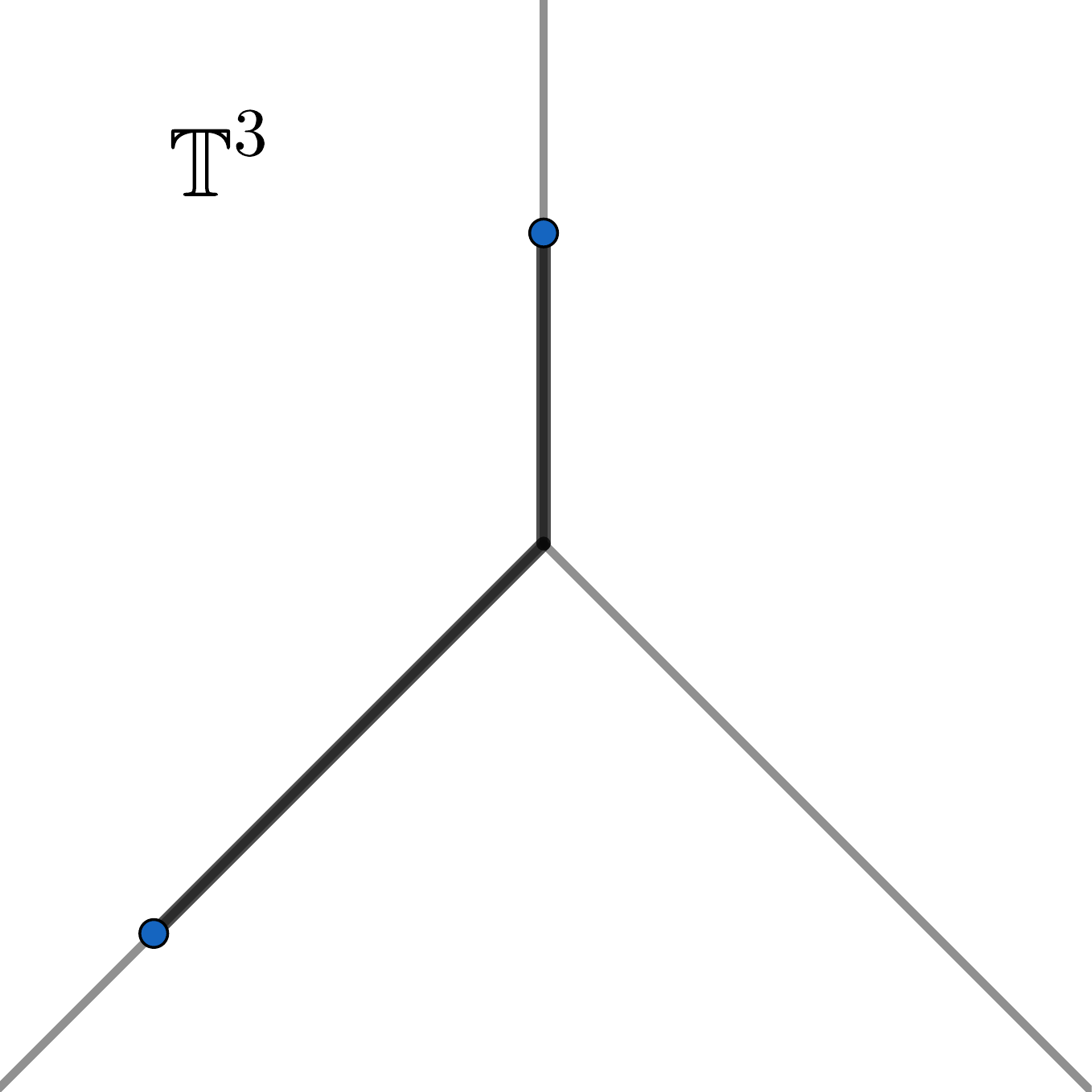}%
\end{minipage}%
\end{minipage}%
}\caption{\label{fig:spaces}Left: Relationship between different complete and
connected metric spaces, with a few commonly investigated metric spaces
shown in dots. Upper right: The two-dimensional unit sphere $\protect\bbS^{2}$
is a Riemannian manifold that is smooth at all points. The distance
between two points (solid dots) is given by the length of the segment
of a great circle connecting them. Lower right: The space of 3-spider
consisting of three Euclidean positive axes issuing from the origin.
This is a geodesic space but not a Riemannian manifold due to the
singularity at the origin. A geodesic connecting two points on different
branches is highlighted, and the distance between points is the length
of the geodesic.}
\end{figure}

The left panel in \ref{fig:spaces} illustrates the relationship between
different types of complete and connected metric spaces and highlights
four common examples. The unit sphere $\bbS^{2}$ in $\bbR^{3}$ is
a Riemannian manifold where a geodesic is a segment of a great circle
(upper right, \ref{fig:spaces}), and the 3-spider that models trees
with three leaves (lower right, \ref{fig:spaces}) is an example of
a geodesic space that is not a Riemannian manifold, since the origin
is degenerate and does not have a neighborhood resembling a real interval. 
%The precise definition of different data spaces are given
%in \ref{sec:depthInMetricSpace} and the Appendix. 

\begin{comment}
While our metric halfspace depth applies to any metric spaces, an
arbitrary metric may not provide meaningful information on the dataset.
(TALK ABOUT THE DISCRETE TOPOLOGY?) Geometric information about the
dataset should be rooted in the metric used. Oftentimes, data lie
in some special object space that come with an intrinsic metric that
specifies a nice geometrical structure. 
\end{comment}

{
If $\cM$ is an unbounded Riemannian manifold such as the space of symmetric positive definite matrices and the hyperbolic space, 
depth notions could alternatively be developed through mapping data on $\cM$ to a tangent space $T_x\cM$ through the inverse exponential map $\exp_x^{-1}: \cM \rightarrow T_x\cM$, and then employing Euclidean depth notions such as Tukey's depth on the linear tangent space.
However, this tangent space approach has limited applicability since it relies on the exponential map  being injective, which is violated on bounded spaces such as the unit sphere $\bbS^k$ and the rotational group $\SO(k)$; 
even if this approach can be applied, it is in general not possible to fully preserve the data geometry reflected by the distance metric while working on the the linear tangent space $T_x\cM$; and the base point $x$ must be chosen.
In contrast, our metric halfspace depth is well-defined and geometry preserving on any metric space.
}%
\subsection{Examples: Metric Halfspace Depth in Common Spaces\label{ssec:examplesMetricSpaces}}

\begin{figure}[h!]
\centering
\resizebox{.7\textwidth}{!}{%
\begin{minipage}[t]{\textwidth}
\includegraphics[height=0.36\textheight]{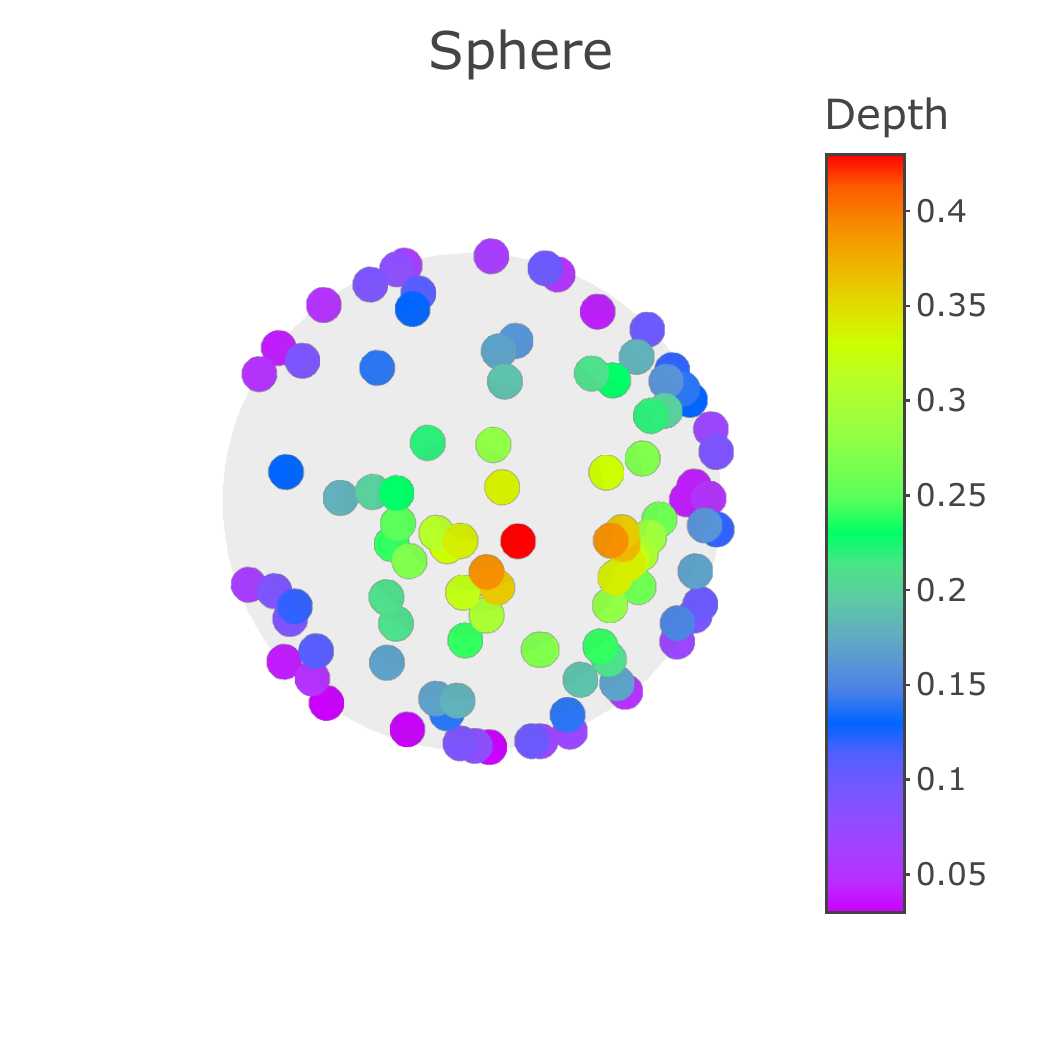}\hfill{}\includegraphics[width=0.59\textwidth]{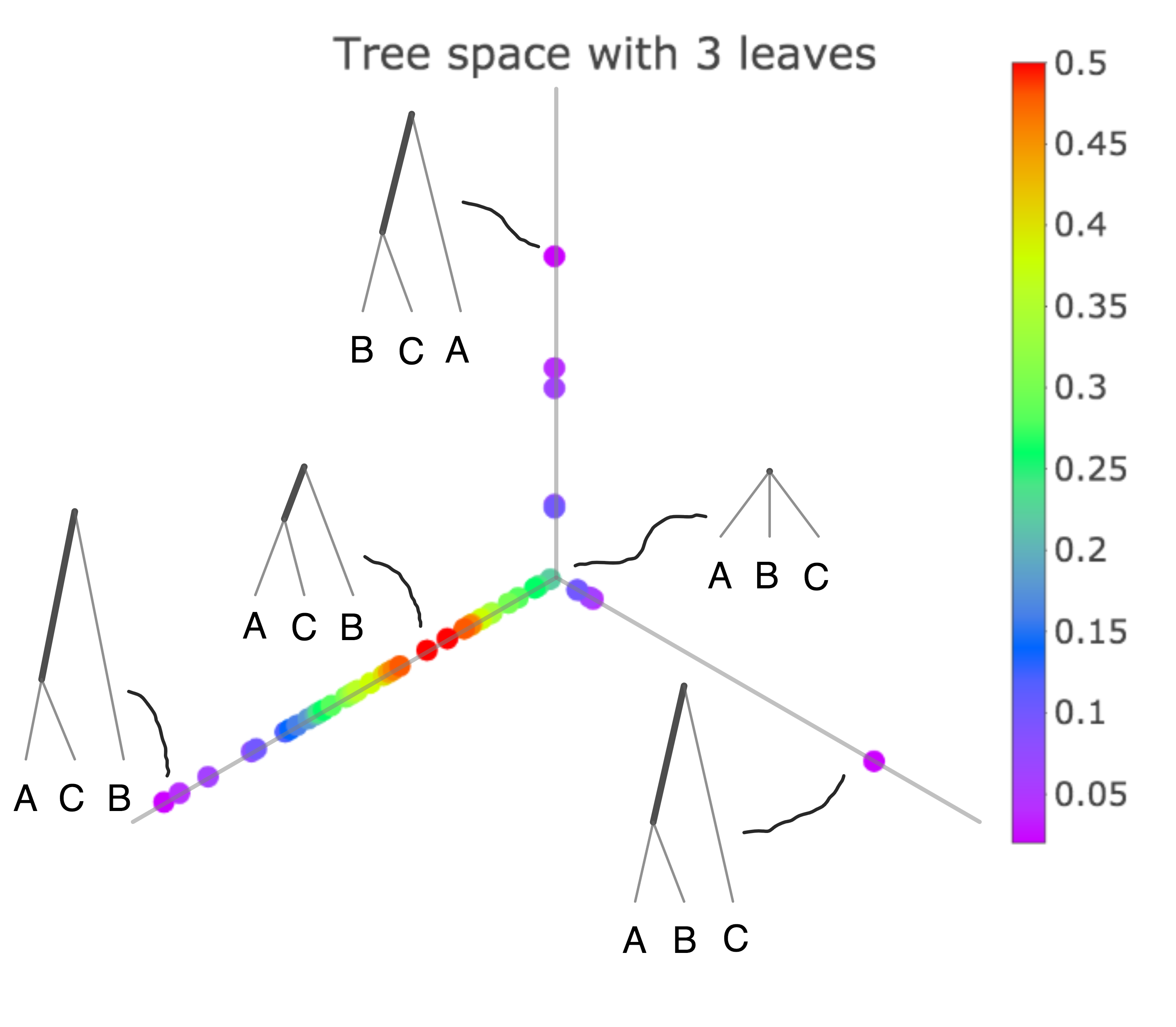}

\centering \includegraphics[height=0.35\textheight]{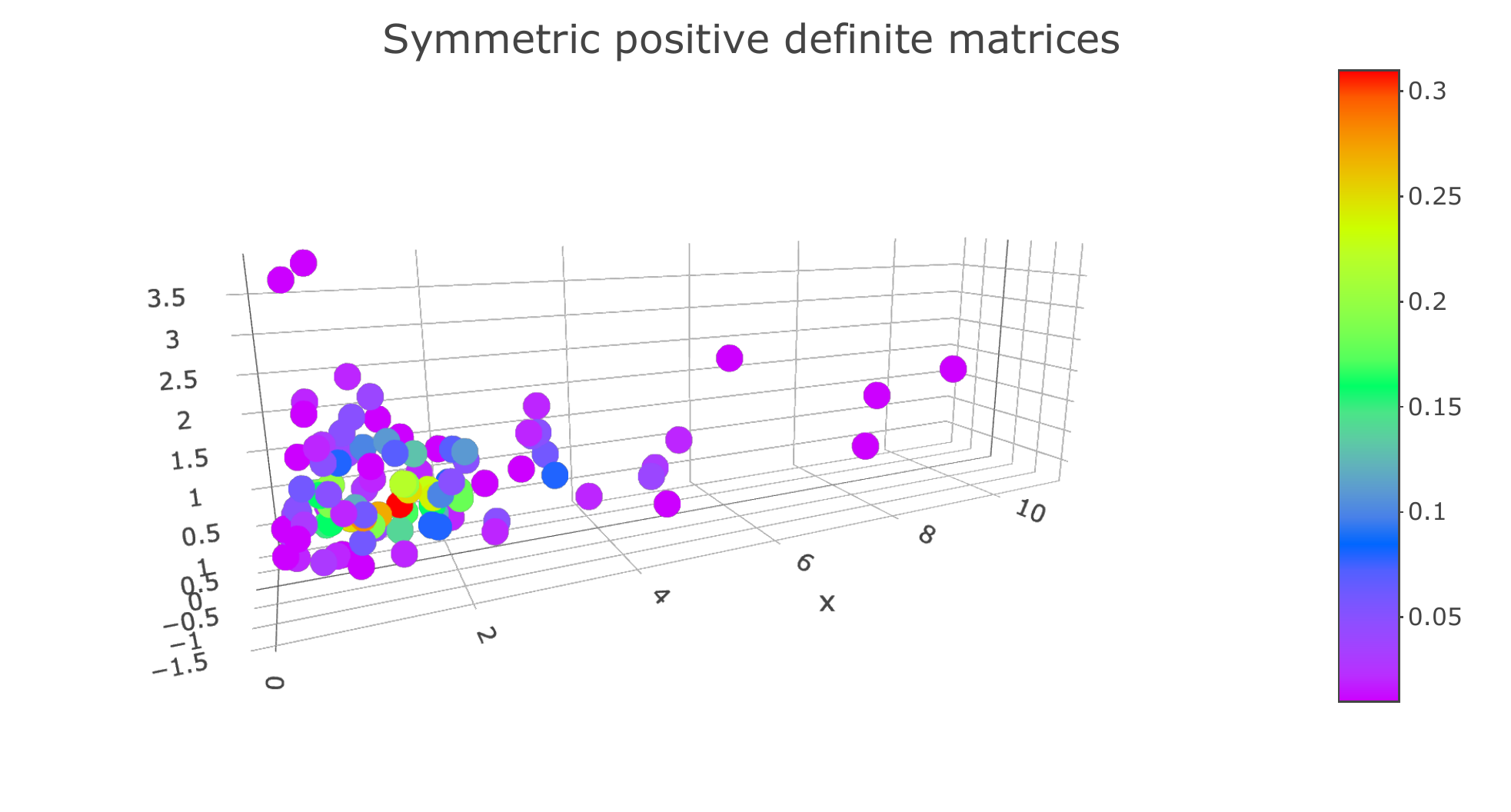}
\end{minipage}%
}

\caption{\label{fig:depthSpaces}Illustration of the proposed metric halfspace
depth of 100 data points generated on different manifolds. Upper left
panel, data followed wrapped normal distribution on the sphere $\protect\cM=\protect\bbS^{2}$.
Upper right panel, data followed a normal distribution centered at
a tree that has leaf B branched out first (on the axis pointing to
eight o'clock). Each dot represents a tree and five trees are
drawn for illustration. Each axis corresponds to a different tree topology and the location on the axis corresponds to the length of the interior edge (bolded). 
The origin corresponds to the star tree which trifurcates at the root node and has a degenerate interior edge. 
Lower panel, $2\times2$ symmetric positive definite matrices $(x~y;\,y~z)$ were generated from a log-normal distribution.
% and analyzed with the affine invariant geometry. 
%More details of the data generating models are included in the Supplemental Materials.
}
\end{figure}

In what follows, we provide examples of commonly investigated data spaces and illustrate  metric halfspace depth in these spaces.
\ref{exa:euclidean}--\ref{exa:spd} concern Riemannian
manifolds,
% $\bbR^{m}$, $\bbS^{m}$, and $\SPD(k)$, 
and \ref{exa:tree}
considers the tree space as a geodesic space that is not a Riemannian
manifold. 
{Depth values in the examples are calculated using the approximation algorithm described in \ref{ssec:approximationAlgorithms}.}
More details about the setups can be found in \ref{assec:setupsDemo} in the Supplemental Materials.%
\begin{example}[Euclidean space]
\label{exa:euclidean}When data lie in the Euclidean space $\cM=\bbR^{m}$, the proposed metric halfspace depth coincides with the Tukey's depth. With norm $\norm x=(x^{\transpose}x)^{1/2}$,
distance $d(x_{1},x_{2})=\norm{x_{1}-x_{2}}$, and Euclidean
halfspace $H'_{x_{0},v}=\{y\in\bbR^{m}\mid(y-x_{0})^{\transpose}v\le0\}$,
the classical Tukey's depth is, 
\begin{equation}
D_{\Tukey}(x)=\inf P_X(H'), \quad x\in\bbR^m, \label{eq:Tukey}
\end{equation}
where the infimum is taken over all Euclidean halfspaces $H'$ containing $x$. 
Any metric halfspace $H=\Honetwo\in\cH$ coincides with
a Euclidean halfspace $H'_{(x_{0},v)}=\{y\in\bbR^{m}\mid(y-x_{0})^{\transpose}v\le0\}$
with $x_{0}=(x_{1}+x_{2})/2$ and $v=(x_{2}-x_{1})/\norm{x_{2}-x_{1}}$
if $x_{1}\ne x_{2}$; visa versa, each Euclidean halfspace can 
be expressed as a metric halfspace. 
Thus, the metric halfspace depth coincides with Tukey's depth because the infimums are
taken over an identical set, noting that $P_X(\Honetwo)=1$ if $x_{1}=x_{2}$
which does not influence the infimum for the metric halfspace depth. 

Tukey's depth in the Euclidean space satisfies all four axiomatic
properties of a depth function introduced in \cite{zuo:00-1}, is a continuous function
of the depth location \citep{mass:04} and can be consistently
estimated by its sample version \citep{mass:04}; 
{
moreover, the deepest
point, i.e. the Tukey's median, has a high breakdown point \citep{dono:92,liu:17} and can also be consistently estimated \citep{bai:99,chen:18-3,zuo:20}.
}%
We will show in \ref{sec:properties} that many of these properties
generalize on geodesic spaces.
\end{example}
\begin{example}[Spheres]
\label{exa:sphere} The $m$-dimensional unit sphere $\bbS^{m}=\{x\in\bbR^{m+1}\mid x^{\transpose}x=1\}\subset\bbR^{m+1}$
is a Riemannian manifold. % used to model directional and spherical data.
The distance between $x,y\in\bbS^{m}$ is the great arc distance $d(x,y)=\arccos(x^{T}y)$.
\begin{comment}
A naive application of the classical Tukey's depth to spherical data
by ignoring the manifold constraint and treating data in the ambient
Euclidean space results in zero depth at all locations. This calls
for proper generalization of the Tukey's depth for non-Euclidean data. 
\end{comment}
The metric halfspace depth specializes to the angular Tukey's depth
for spherical data considered by \citet{smal:87,liu:92}, where the latter is defined as the least probability measure of any hemisphere covering $x$. 
This is because a metric halfspace $\Honetwo=\{x\in\bbS^{m}\mid x^{\transpose}(x_{2}-x_{1})\le0\}$ is a closed hemisphere in this context.
% and thus the metric halfspace depth
%coincides with the angular Tukey's depth, where the angular Tukey's
%depth at $x\in\bbS^{m}$ is .

An example of the metric halfspace depth applied to data on $\bbS^{2}$
is shown in the upper left panel of \ref{fig:depthSpaces},
{where data were generated according to the wrapped normal distribution with isotropic variance $1/2$; the setup is described in \autoref{assec:setupsDemo}.}
%by projecting normally distributed data on a tangent space to the sphere. 
The depth values follow a center-outward pattern,
monotonically decreasing from the deepest point near the center of symmetry.%
\begin{comment}
; a rigorous proof will be shown in \ref{sec:properties}
\end{comment}
{} The deepest point meaningfully characterizes a representative point
well-encompassed by the point cloud, and the points with the lowest depth
all lie on the peripheral.
\end{example}
\begin{example}[Symmetric positive definite matrices]
\label{exa:spd} Let $\cM=\SPD(k)$ be the manifold of $k\times k$
symmetric positive definite (SPD) matrices. This matrix manifold has
seen wide application in modeling brain connectivity matrices \citep{dai:20}
and diffusion tensors \citep{penn:06}. Endowed with the affine-invariant
geometry \citep{penn:06}, the geodesic distance on $\cM$ is defined as $d(P,Q)=\normF{\logm(P^{-1/2}QP^{-1/2})}$
for $P,Q\in\cM$, where $\normF{\cdot}$ is the Frobenius norm, $\logm$
is the matrix logarithm, and $P^{-1/2}$ is the inverse of the symmetric
positive definite square root $P^{1/2}$ of $P$. The geometry is
invariant under affine transformations in the sense that $d(APA^{\transpose},AQA^{\transpose})=d(P,Q)$
for any invertible matrix $A$ and thus have been widely adopted in
applications. 
{
As the data space is non-Euclidean with a complex
geometry, the halfspaces in general have rather complex shapes. An example of a halfspace $\Honetwo$ is shown in \ref{fig:spdHalfspace}. 
Here we evaluate the depth of an SPD matrix with respect to a sample of SPD matrices as the data units, which is different from the scenario considered for scatter depth \citep{chen:18-3,pain:18} where the depth of an SPD matrix is evaluated with respect to Euclidean data units for estimating the covariance matrix.
}%

Illustrated for $\cM=\SPD(2)$, the lower panel of \ref{fig:depthSpaces} displays non-isotropic log-normal matrix data points that are colored according to the proposed metric halfspace depth. 
Each point represents the lower diagonal values of an SPD matrix $\left(\begin{smallmatrix}x & y\\
y & z
\end{smallmatrix}\right)$. 
The proposed depth produces reasonable results by showing a center-outward profile analogous to Tukey's depth in the Euclidean space. 
The deepest point in red is tightly surrounded by data points with gradually decreasing depth values, and the deepest point is not heavily drawn by data points with large values in the diagonal elements $x$ and $z$. 
The peripheral points all have the least depth. 
\end{example}
\begin{example}[BHV space of phylogenetic trees]
\label{exa:tree}We model phylogenetic trees in the Billera--Holmes--Vogtmann (BHV) tree space \citep{bill:01},
a widely investigated geodesic space with nice geometry.
% and fast algorithms for finding the geodesics \citep{owen:11}. 
Let $\cM=\bbT^{k}$ denote the space of rooted phylogenetic trees with $k$ leaves endowed with the BHV geometry \citep{bill:01}, where a brief summary for the BHV geometry and the associated metric halfspaces is included in \ref{assec:tree}. 

We illustrate here the geometry of the simplest tree space $\bbT^{3}$ with three
leaves and one interior edges. 
The \emph{topology} of the tree is the way leaves and interior nodes are connected. 
There are three bifurcating tree topologies respectively corresponding to which of leaf A, B, and C branches out first, and a star tree topology with a degenerate interior edge. 
Tree space $\bbT^{3}$ is represented by the 3-spider $(\bbRnn\times\{1,2,3\})/\sim$, formed by three
rays identified at the origin $o$.
Coordinate $(a, j)$ represents a point (tree) lying on the $j$th leg of the 3-spider at a distance $a$ from the origin; we refer to this representation of the trees as the (radius, branch)-coordinate.
The equivalence relationship $\sim$ is defined by $(a_{1},j_{1})\sim(a_{2},j_{2})$ if and only if $(a_1,j_1)=(a_2,j_2)$ for $a_1 > 0$ and for $a_1 = a_2=0$.
The three legs of the spider correspond to
three different bifurcating tree topologies, and the position of a 
point on a leg corresponds to the length of the interior edge, as
illustrated in \ref{fig:depthSpaces}. 
The geodesic between two points on the same branch is the line segment connecting them; 
analogously, the geodesic between two points on different branches consists of the line segments connecting each to the origin. 
Thus,
the distance between two points $x,y$ on the 3-spider is the Euclidean
distance if they are on the same branch, and $d(x,o)+d(o,y)$ if they
are on different branches (see lower right panel, \ref{fig:spaces}).
%Halfspaces in $\bbT^{3}$ takes one of three possible forms displayed in \ref{fig:treeHalfspace}. 
%A brief discussion of the geometry for the general case $\bbT^{k}$ is included in the Appendix, while for full details we refer to \citet{bill:01}. 

An illustration of the metric halfspace depth for trees with three leaves on $\bbT^{3}$ is shown in \ref{fig:depthSpaces}.
%, where the data included trees with all three fully-resolved topology.
%Each axis on the 3-spider corresponds to a different tree topology, depending on which of leaf A, B, or C branches out first, and the location on each axis corresponds to the length of the interior edge.
The trees were generated according to a normal distribution centered at a tree with leaf B branched out first (on the axis pointing to 8 o'clock). The proposed metric halfspace depth assigned the largest value for trees around the center, and the depth values gradually and monotonically decreased as data moved away from the center. The
depths of the most peripheral trees on each axis received the lowest
depths. 
A small number of trees had either leaf A or C branching out first,
and these trees were all assigned low depths. 
\end{example}

\section{Theoretical Properties\label{sec:properties}}

\subsection{Desirable Depth Properties \label{ssec:axiomaticProperties}}

\begin{comment}
For finite dimensional Riemannian manifolds the theory may be obtained
through empirical process theory. What we need would be the collection
of halfspaces not to be too large. For the infinite dimensional functional
data, the numerical results work, but I don't have an idea what (\ref{eq:DsampApprox})
is approaching -- definitely not simply the Tukey's halfspace depth,
since it is 0 most of the time \citep{dutt:11}. 
\end{comment}

For a data depth notion to quantify reasonably how representative data points are within a distribution or sample, and define a center-outward ordering, \citet{zuo:00-1} postulated four properties that the depth function should satisfy when analyzing data in a Euclidean space, namely (a) \emph{Affine
invariance}, i.e.\ the depth of a point is invariant to affine transformations; (b) \emph{Vanishing at infinity}, namely the depth should
approach 0 as one moves away from the center of the data; (c) \emph{Maximality at the symmetric center}, namely if there is a  ``center'', such as a point of symmetry, in the data, then the depth achieves its maximum at this center; and (d) \emph{Center-outward monotonicity}, i.e.\ depth values gradually decrease
as one moves away from the deepest point. These properties are satisfied
by the classical Tukey's depth \citep{tuke:75}, and we will show
that to a great extend they are also satisfied by the proposed metric
halfspace depth.

The four depth properties are satisfied for the proposed metric halfspace
depth under regularity conditions as detailed in the next theorem.
To state these properties in a general metric space, due to the lack
of a vector space structure, we need to address the lack of affine
data transformation and introduce an invariance property, a notion of data symmetry, and monotonicity. 

For an invariance property, let $f$ be a transformation from $(\cM,d)$
to another metric space $(\cN,e)$. For any $y\in\cN$, let $\cH_{y,e}=\{\Hyote\mid y\in\Hyote,\,y_{1},y_{2}\in\cN\}$
be the collection of halfspaces $\Hyote=\{z\in\cN\mid e(z,y_{1})\le e(z,y_{2})\}\subset\cN$
containing $y$. We say that $f$ is \emph{halfspace preserving at
$x\in\cM$ with respect to $(\cM,d)$ and $(\cN,e)$, }or simply \emph{halfspace preserving} at $x$ if $\cH_{f(x),e}=f(\cH_{x})\coloneqq\{f(H)\mid H\in\cH_{x}\}$, in which case the collection of halfspaces containing $x$ is preserved by $f$. 
We say that $X$ is \emph{halfspace symmetric} about $\theta\in\cM$
if $P(X\in H)\ge1/2$ for all halfspace $H$ containing $\theta$, {extending the same notion defined in the Euclidean space by \cite{zuo:00-1}}. 
To define monotonicity on a metric space, we restrict attention to
geodesic spaces, where monotonicity of the depth function can be investigated
along geodesics leaving from the deepest point. 

For theory development, we require $(\cM,d)$ to be a connected complete separable metric space. 
For a subset $S\subset\cM$, let $S^{\circ}$, $\overline{S}$, $\partial S$,
and $S^{\comp}$ denote the interior, closure, boundary, and complement
of $S$, respectively. 
%recall that the halfspace $\Honetwo$ anchored at $x_{1},x_{2}\in\cM$ is defined in (\ref{eq:halfspace}), and we 
Proofs for the theoretical results and additional analytical properties of the halfspaces are included in \autoref{asec:proofs} and \autoref{assec:propertiesHalfspaces}, respectively. 
\begin{thm}
\label{thm:fourPostulates}The metric halfspace depth $D(\cdot)$ satisfies
the following properties.
\begin{enumerate}[label=(\alph*)]
\item \label{enu:invariance}\emph{(Transformation invariance)} Let $f:\cM\rightarrow\cN$
be a bijective measurable map between metric spaces $(\cM,d)$ and
$(\cN,e)$, and $D_{f}(y)=\inf_{H'\in\cH_{y,e}}P_{f(X)}(H')$ denote
the depth at $y\in\cN$ with respect to the pushforward measure $P_{f(X)}=P_X\circ f^{-1}$ on $\cN$. If $f$ is halfspace preserving at $x\in\cM$, then $D(x)=D_{f}(f(x))$. 
\item \label{enu:vanish}\emph{(Vanishing at infinity)} Let $o\in\cM$ be
an arbitrary point. Then $\sup_{x:d(o,x)>L}D(x)\tozero$ as $L\toinf$,
taking the convention here that the supremum over an empty set is
0. 
\item \label{enu:center}\emph{(Maximality at the symmetrical center)} If
$X$ is halfspace symmetric about a unique center $\theta$, then
$\theta$ is the unique deepest point, i.e., $\theta=\argmax_{x\in\cM}D(x)$. 
\item \label{enu:monotonicity}\emph{(Center-outward monotonicity)} Suppose $\cM$ is a geodesic space. Let $\theta\in\cM$ be
a deepest point, $x\in\cM$, and $\gamma:[0,1]\rightarrow\cM$ a geodesic
joining $\theta$ to $x$. If any halfspace $\Honetwo$ of $\cM$
that has a nonempty intersect with $\gamma([0,1])$ contains at least
one of $x$ and $\theta$, then $D(x)\le D(\gamma(t))$ holds for
$t\in[0,1]$. 
\end{enumerate}
\end{thm}
\ref{thm:fourPostulates}\ref{enu:invariance} states that the metric
halfspace depth is invariant to transformation $f$ that preserves halfspaces.
It is immediate that affine transformations and rotations are halfspace
preserving, respectively, between Euclidean spaces and between spheres of the same dimension at all $x\in\cM$. 
Thus, this result implies the transformation invariance properties of Tukey's depth \citep{dono:92} and angular Tukey's depth \citep{liu:92}. 
More generally, a map $f$ is halfspace preserving
at $x$ if it preserves the order of distances at $x$, i.e., for $x,x_{1},x_{2}\in\cM$, $d(x_{1},x)\le d(x_{2},x)$ if and only if $e(f(x_{1}),f(x))\le e(f(x_{2}),f(x))$. 
This is clearly satisfied if $f$ is an isometry, i.e., $d(x,y)=e(f(x),f(y))$ for all $x,y\in\cM$. 

The depth follows a center-outward tendency. In a space where ``infinite'' is well-defined, \ref{thm:fourPostulates}\ref{enu:vanish} states that the depth of a point vanishes as the point moves towards infinity.
Therefore,
the peripheral data points will have a small depth.
\ref{thm:fourPostulates}\ref{enu:center} states that if the data
distribution is halfspace symmetric about a unique center $\theta$, then the halfspace depth is maximized at this center $\theta$. 
We consider halfspace
symmetry to define data symmetry on a general metric space, which
does not require the space $\cM$ itself to be symmetric, thereby generalizing beyond the Euclidean space and spheres \citep{liu:92}. 
%A special case of the halfspace symmetry condition in the Euclidean space has been considered by 
In the Euclidean space, \citet{zuo:00-1} showed that halfspace symmetric is weaker than alternative symmetry notions such as \emph{centrally symmetric}, i.e. $X-\theta$ and $\theta-X$ equal in distribution, and \emph{angularly symmetric}, which requires $(X-\theta)/\norm{X-\theta}$ to be centrally symmetric. 

Between the deepest $\theta$ and an arbitrary
location $x$, \ref{thm:fourPostulates}\ref{enu:monotonicity} states
that the metric halfspace depth is non-increasing along geodesics
leaving from $\theta$ if the metric space satisfies a geometric condition.
The geometric condition requires that the halfspaces in $\cM$ are not overly rich so they will not single out points on the geodesic connecting $\theta$ and $x$ while excluding the endpoints. 
This condition is satisfied by
the model spaces, namely the Euclidean space, sphere, and hyperbolic space, as stated in \ref{prop:HSDoNotSingleOut}. 
\begin{comment}
In general, this condition is not necessary for the monotonicity of
depth, since near monotonicity of depth with respect to the deepest
point is often observed in data applications even in metric spaces
with rich halfspaces that violate this condition, such as the tree
space $\bbT^{k}$.
\end{comment}

\begin{prop}
\label{prop:HSDoNotSingleOut}Let $\cM$ be one of the $m$-dimensional
model spaces, namely, the Euclidean space $\bbR^{m}$, unit sphere
$\bbS^{m}$, or hyperbolic space $\bbH^{m}$, and $\gamma:[0,1]\rightarrow\cM$
be a geodesic joining $\theta$ to $x$. Then any halfspace $H\subset\cM$
with a nonempty intersect with $\gamma([0,1])$ contains at least one
of $\theta$ and $x$.
\end{prop}
%

% \subsection{Semi-continuity\label{ssec:semicontinuity}}

We next show the upper semi-continuity of the depth function $D(\cdot)$
and the compactness and nestedness of the depth regions $D^{\alpha}\coloneqq\{x\in\cM\mid D(x)\ge\alpha\}$, $\alpha>0$.
Define $P_H: \cM\times\cM \rightarrow \bbR$ as $P_H(x_{1},x_{2})=P_X(\Honetwo)$ and let $\Eonetwo=\{x\in\cM\mid d(x,x_{1})=d(x,x_{2})\}$ be the equidistance set anchored at $x_{1},x_{2}\in\cM$. 
A metric space is \emph{locally compact} if every point has a compact neighborhood. All finite-dimensional
manifolds and BHV tree spaces are locally compact. 

%The following condition
%is needed which states that equal distance sets $\Eonetwo$ have zero
%probability. 
%
%\begin{enumerate}[label=(P)]
%\item \label{a:EProb0} $P(\Eonetwo)=0$ for all $x_{1}\ne x_{2}\in\cM$.
%\end{enumerate}
%
%
\begin{prop}
\label{prop:semicontinuity}Suppose that $\cM$ is a complete and
locally compact geodesic space.
\begin{enumerate}[label=(\alph*)]
\item \label{enu:PSemiCont} $P_H(\cdot,\cdot)$ is upper-semi continuous. If further $P_X(\Eonetwo)=0$ for all $x_{1}\ne x_{2}\in\cM$, then $P_H(\cdot, \cdot)$ is continuous. 
\item \label{enu:DxSemiCont} $D(x)$ is upper semi-continuous. 
\item \label{enu:DxRegion}$D^{\alpha}$ is nested, i.e. $D^{\alpha_{1}}\subset D^{\alpha_{2}}$
for $\alpha_{1}\ge\alpha_{2}$, and $D^{\alpha}$ is compact for $\alpha>0$.
\end{enumerate}
\end{prop}
%The same proof steps show that if \ref{a:EProb0} does not hold, then
%the halfspace probability is upper semi-continuous. 
The additional condition in \ref{prop:semicontinuity}\ref{enu:PSemiCont} is satisfied 
%The equal-distance sets $\Eonetwo$, $x_1\ne x_2$ have probability zero 
if $\cM$ is a Riemannian manifold and $X$ has a density w.r.t. the Riemannian volume measure \citep{lee:18}; for example, this is satisfied if $\cM$ is the unit sphere and $X$ follows a warped normal distribution. 

\begin{comment}
%
\begin{prop}
\label{prop:depth-1}The following holds in a complete connected metric
space $\cM$:
\begin{enumerate}[label=(\alph*)]
\item \label{enu:PSemiCont-1} $H(\cdot,\cdot)$ is upper semi-continuous.
If \ref{a:EProb0} further holds, then $H(\cdot,\cdot)$ is continuous.
\item \label{enu:DxSemiCont-1}If \ref{a:moveHS} holds, then $D(x)$ is
upper semi-continuous.
\item \label{enu:DxRegion-1}$D^{\alpha}$ is nested, i.e. $D^{\alpha_{1}}\subset D^{\alpha_{2}}$
if $\alpha_{1}\ge\alpha_{2}$. If \ref{a:moveHS} holds, then $D^{\alpha}$
is compact for $\alpha>0$.
\end{enumerate}
\end{prop}
\end{comment}

\subsection{Convergence of the Depth Function and Deepest Point\label{ssec:convergence}}

Next, we show that the metric halfspace depth can be estimated consistently
by its sample version uniformly over all locations by making use of
empirical process theory. Let $L_{2}(Q)$ be the $L_{2}$-norm of
measurable functions with respect to probability measure $Q$ on the
sigma-algebra of $\cM$, so $L_{2}(Q)(f)=\{\int f(x)^{2}dQ(x)\}^{1/2}$.
For a set of measurable functions $\cF$, the \emph{covering number}
$N(\epsilon,\cF,L_{2}(P))$ is the minimal number of balls in $L_{2}(P)$
with radius $\epsilon$ required to cover $\cF$. %
\begin{comment}
The \emph{entropy} is the logarithm of the covering number. 
\end{comment}
The \emph{bracketing number} $\Nbrac(\epsilon,\cF,L_{2}(Q))$ is the
minimal number of $\epsilon$-brackets required to cover $\cF$. An
$\epsilon$\emph{-bracket} $[l,u]$ is the set of functions $f$ with
$l\le f\le u$, given two functions $l$ and $u$ with $\norm{u-l}_{L_{2}(Q)}<\epsilon$.
The covering and bracket numbers for a collection of measurable sets are by convention those of the corresponding collection of indicator functions. 
Either one of the following conditions is needed for the convergence results. 
\begin{enumerate}[label=(N\arabic*)]
\item  \label{a:uniformEntropy}$\int_{0}^{\infty}\sup_{Q}[\log N(\epsilon,\cH,L_{2}(Q))]^{1/2}d\epsilon<\infty$,
where the supremum is taken over all finite discrete probability measures
$Q$ and $\cH$ is the set of metric halfspaces.
\item \label{a:bracketIntegral}$\int_{0}^{\infty}[\log\Nbrac(\epsilon,\cH,L_{2}(P))]^{1/2}d\epsilon<\infty$.
\end{enumerate}
\begin{thm} %[$D_{n}(x)$ is $n^{1/2}$-convergence]
\label{thm:DnRate} If either \ref{a:uniformEntropy} or \ref{a:bracketIntegral}
holds, then 
\[
{E \supxM\left|D_{n}(x)-D(x)\right|=O(n^{-1/2}).}
\]
\end{thm}
Condition \ref{a:uniformEntropy} and \ref{a:bracketIntegral} are
common entropy/bracketing integral conditions imposed on the complexity of the collection of halfspaces in order to guarantee convergence of the empirical process.
Condition \ref{a:uniformEntropy} holds if the Vapnik--Chervonenkis
(VC) dimension of $\cH$ is finite \citep[Theorem 2.6.4,][]{van:96}.
Let $\cC$ be a collection of subsets of $\cM$. We say that $\cC$
shatters a finite subset $F=\{x_{1},\dots,x_{n}\}\subset\cM$ if $\cC\cap F\coloneqq\{C\cap F\mid C\in\cC\}$
is the collection of all subsets of $F$. 
{
The \emph{Vapnik--Chervonenkis
(VC) dimension} of $\cC$ is the smallest $n$ for which no set of size $n$ is shattered by $\cC$, formally defined by
\begin{align*}
\VC(\cC) & = \inf\{ n \mid \max_{x_1,\dots,x_n} \Delta_n (\cC, x_1,\dots,x_n) < 2^n \},
\end{align*}
where $\Delta_n(\cC,x_1,\dots,x_n) = \left|\{ C \cap \{x_1,\dots,x_n\}\mid c\in\cC \} \right|$ is the number of subsets of $\{x_1,\dots,x_n\}$ picked out by $\cC$.
}%
%$\VC(\cC)=\max\{\left|F\right|\mid F\text{ is shattered by }\cC\}$,
%the largest cardinality of sets shattered by $\cC$. 
It is well known
that the VC dimension of halfspaces in the Euclidean space $\bbR^{m}$
is $m+2$ \citep{weno:81}. 
Theory on the VC dimensions of subsets of a Riemannian manifold \citep{nara:09} or of a general metric space has been highly limited. 
That said, since the collection of halfspaces $\cH$ is indexed by two points on the metric space, it may be reasonable to expect $\VC(\cH)$ to be finite if the geometry of $\cM$ is regular enough. 
We establish the boundedness of VC dimensions
for the collections of halfspaces on the sphere $\bbS^{m}$ and the
space $\bbT^{3}$ of phylogenetic trees with 3 leaves.
\begin{prop}
\label{prop:VC}The following holds: 
\begin{enumerate}[label=(\alph*)]
\item \label{enu:sphereVC}On an $m$-dimensional sphere $\cM=\bbS^{m}$,
${\VC(\cH)\le m+3}$. %
\begin{comment}
\label{enu:torusVC}On the flat torus $\cM=\bbS^{1}\times\bbS^{1}$,
$\VC(\cH)<\infty$. 
\end{comment}
\item \label{enu:treeVC}On the space of phylogenetic trees $\cM=\bbT^{3}$
with $3$ leaves, ${\VC(\cH)=4}$. %
\begin{comment}
On the space of phylogenetic trees $\cM=\bbT^{m}$ with $m$ leaves,
$\VC(\cH)<\infty$. 
\end{comment}
\end{enumerate}
\end{prop}
\noindent By \ref{thm:DnRate}, \ref{prop:VC} implies $n^{1/2}$-convergence for the empirical metric halfspace depth on these spaces.

A deepest point w.r.t.\ the sample is a consistent estimator of the population deepest point by $M$-estimation theory. 
{
\begin{prop}
\label{prop:thetaConsistent}
Suppose that $\cM$ is a complete and
locally compact geodesic space, $D(\cdot)$ has a unique maximum $\theta=\argmax_{x\in\cM}D(x)$,
and the conditions of \ref{thm:DnRate} hold. Let $\theta_{n}$ be
an arbitrary point in the deepest set $S_n \coloneqq \argmax_{x\in\cM}D_{n}(x)$.
Then
\[
d(\theta_{n}, \theta) \tozero\quadas
\]
as $\ntoinf$. 
\end{prop}

By \autoref{thm:fourPostulates}\autoref{enu:invariance}, the deepest set $S_n$ is invariant to halfspace preserving transformations. 
In the asymptotic limit, the deepest set shrinks to the population deepest point if the latter is unique, so any sample deepest point is near-invariant. 
}%

\subsection{Robustness\label{ssec:robustness}}

{
A depth median is defined as an estimator $T(\cdot)$ that takes a point cloud $\cZ = \{z_1,\dots,z_n\}$ on $\cM$ to a choice of point $T(\cZ) \in \argmax_{x\in\cM} D(x;P_\cZ)$ in the deepest set w.r.t. metric halfspace depth, where $P_\cZ$ is the empirical measure placing equal point mass on each point in $\cZ$. 
Given a sample, a depth median yields a (unique) point as output, but there exists potentially more than one depth medians (as estimators) in general if the deepest set is non-singleton. 
The depth median of a point cloud is interpreted as the most representative point of the data and can be used as a location descriptor/estimator. 
In the Euclidean case, Tukey's depth median is a generalization of the classical median on the real line and satisfies robustness \citep{dono:92} and asymptotic properties \citep{mass:04}. 
%For our metric halfspace depth, we term the depth median as \emph{metric halfspace median}, and here we analyze its breakdown property. 

The breakdown property of the \emph{metric halfspace median}, which is the depth median based on our metric halfspace depth, is analyzed next.
%We analyze the robustness of the depth median w.r.t. the metric halfspace depth, which we term as metric halfspace median, by analyzing its breakdown property. 
Intuitively, the breakdown point is the smallest fraction of contamination that brings an estimator to infinity. 
Formally, let $\cXn = \{X_1,\dots,X_n\}$ be a sample of $n$ observations and $\cYl = \{y_1,\dots,y_{l}\}$ be $l$ contamination points. 
The \emph{breakdown point} $\epsilon^{*}$ of a metric halfspace median $T(\cdot)$ in a sample $\cXn$ is the smallest fraction of contamination to bring the estimate in the contaminated sample arbitrarily far away from that of the uncontaminated sample. 
The finite-sample (additional) \emph{breakdown point} is defined as
\[
\epsilon^{*}= \epsilon^*(T;\cXn) \coloneqq 
\min_{l}\left\{ \frac{l}{n+l}\left|
\sup_{\cYl} d(T(\cXn), T(\cXn,\cYl) ) = \infty
\right.\right\} ,
\]
where we set $\epsilon^{*}=1$ if the set being minimized is empty. 
The next proposition and its corollary analyze the finite-sample and asymptotic behavior of the breakdown point, respectively. 
\begin{prop}
\label{prop:breakdownSamp} Let $\cM$ be an arbitrary metric space.
For any metric halfspace median $T(\cdot)$, it holds that
\[
\epsilon^{*}\ge\frac{D_{n}(\theta_{n})}{1+D_{n}(\theta_{n})},
\]
where $\theta_n = T(\cXn)$ is a deepest point w.r.t.\  sample $\cXn$. 
\end{prop}
\begin{cor}
\label{cor:breakdownPop}If \ref{a:uniformEntropy} or \ref{a:bracketIntegral}
holds, then as $\ntoinf$, 
\[
\epsilon^{*}\ge\frac{D(\theta)}{1+D(\theta)}\quadas,
\]
where $\theta \in \argmax_{x\in\cM} D(x)$ is any deepest point w.r.t.\  the distribution $P_X$ of $X$.
\end{cor}
}%
\ref{cor:breakdownPop} implies that the breakdown point for the metric
halfspace median for any halfspace symmetric distribution is at least
$1/3$ regardless of the dimensions, extending the results for Tukey's median in the Euclidean 
case \citep{dono:92}.

\section{Efficient Computation\label{sec:computation}}

\subsection{Approximation Algorithms\label{ssec:approximationAlgorithms}}

In the Euclidean space, exact computation of Tukey's depth and
deepest point are prohibitively slow if the dimension is higher than
3 even with efficient algorithms \citep{dyck:16}.
On a general metric space, the evaluation of the metric halfspace depth as an infimum faces additional difficulty and would require optimization algorithms that adapt to specific manifolds \citep{yang:07}. 
Moreover, the search for the deepest point requires difficult optimization of a discontinuous function $D_{n}(\cdot)$. 
%that we consider, an additional difficulty is taking into account the geometrical structure of $\cM$. 
%For evaluating the depth, general
%optimization algorithms such as the gradient descent \citep{yang:07}
%and quasi-Newton's methods \citep{ring:12} have been developed on
%smooth Riemannian manifolds, but the implementation need to be considered
%on a case-by-case basis for different manifolds; in non-smooth spaces
%like the BHV tree space, specialized algorithms need to be introduced
%\citep{skwe:18}. 
This motivates us to develop fast approximation algorithms
for the metric halfspace depth and deepest point.

Let $\cX=\{X_{1},\dots,X_{n}\}\subset\cM$ be the collection of observations, and also denote $\cA\subset\cM$ as the \emph{anchor set} containing $\left|\cA\right|=n_{\cA}$ anchor points of halfspaces.
We approximate $D_{n}(x)$ w.r.t.\  $\cX$ by taking the infimum over only halfspaces anchored at points in $\cA$.
The proposed metric halfspace depth approximation is
\begin{equation}
\tD_{n}(x) = \tD_n(x;\cA) =\inf_{x_{1}\ne x_{2}\in\cA:\,d(x,x_{1})\le d(x,x_{2})}n^{-1}\sumin\one\{d(X_{i},x_{1})\le d(X_{i},x_{2})\}.  \label{eq:DsampApprox}\\
% & =\inf_{H\in\cH_{x}\cap\cHAA}P_{n}(H).\nonumber 
\end{equation}
The infimum is taken over at most $n_{\cA}(n_{\cA}-1)$ ordered pairs
of anchors. The number of anchors controls the tradeoff between computational
cost and accuracy, in that using a larger number of anchors results in a better approximation but at a higher cost. 
In most applications, the anchor points $\cA$ can be set to the sample points $\cX$, and for improving approximation, one can enlarge the set of anchor points by including ``jiggled'' versions of these points; more information is included in \autoref{assec:jiggle}. 
{
The deepest point is approximated by the in- and out-of-sample points with the largest approximate depth, defined respectively by
\begin{equation}
\ttheta=\argmax_{x\in \cX}\tD_{n}(x), \quad \rtheta = \argmax_{x\in\cM} \tD_n(x).\label{eq:ttheta}
\end{equation}
The in-sample deepest point $\ttheta$ can serve as a good initial value in numerical optimization procedures to search for the out-of-sample deepest $\rtheta$, where the latter is a more accurate approximation of $\htheta$. 
}%

Like their population and sample versions, the approximate depth and
deepest points incorporate the geometry of $\cM$ through the metric
$d$ and thus avoids the choice of a parametrization of the metric space or linearization onto the tangent space, both of which could be ill-defined.
The approximate depth is defined as long as the discrete graph of pairwise geodesic distances is given, and thus the proposed depth is applicable to a wide range of scenarios where the available data are nodes and edges of a graph \citep{smal:97} or where the pairwise geodesics are estimated from a point cloud using a graph-based method \citep{tene:00}. 
Algorithms for computing depth and the deepest point are summarized in \ref{alg:depth} and \ref{alg:deepest}, respectively.

\begin{algorithm}
\caption{\label{alg:depth}Evaluate depth at points in $\protect\cY$ w.r.t.
$\protect\cX$}

\KwData{Random sample $\cX$, depth evaluation points $\cY$, and halfspace anchors $\cA$}
\KwResult{Depths $D(y)$ for $y\in\cY$}

\For{$x_1 \ne x_2 \in \cA$}{ \label{line:allPn}
	$p_{x_1,x_2} \leftarrow P_n(\Honetwo)$ \label{line:Pn}
} \label{line:allPnEnd}

\For{$y \in \cY$}{ \label{line:ycY}
	$Q \leftarrow \emptyset$

	\For{$x_1 \ne x_2 \in \cA$}{
		\If{$d(y,x_1)\le d(y,x_2)$}{
			Add $p_{x_1,x_2}$ to $Q$ 
		}
	}

	$D(y) \leftarrow \min Q$
} \label{line:ycYEnd}
\end{algorithm}%
\begin{algorithm}
\caption{\label{alg:deepest}Locate the deepest point in $\protect\cX$}

\KwData{Random sample $\cX$ and anchor points $\cA$}
\KwResult{Deepest point $\ttheta$}

Obtain $\tD_n(x)$, $x\in\cX$ by invoking \ref{alg:depth} with $\cY=\cX$

$\ttheta \leftarrow \argmin_{x\in\cX} \tD_n(x)$
\end{algorithm}

The complexity of \ref{alg:depth} is $O(n_{\cY}n^{2}+n^{3})$ for  evaluating depth at points in $\cY$ w.r.t.\  sample $\cX$ and anchor points $\cA=\cX$, where $n_\cY=\left| \cY \right|$.
%,  since \ref{line:Pn} takes $O(n)$, \ref{line:allPn}--\ref{line:allPnEnd} takes $O(n^{3})$, and \ref{line:ycY}--\ref{line:ycYEnd} takes $O(n_{\cY}n^{2})$.
%independent of the dimension of the space $\cM$. 
%Calculating the
%distance between a pair of points in an $m$-dimensional space is
%typically $O(m)$ for a Riemannian manifold and $O(m^{4})$ in the
%BHV space $\bbT^{m}$ \citep{owen:11}, so 
The rate of complexity does not have an exponent involving dimension $m$, similar to those of the approximation algorithms \citep[e.g.,][and the references therein]{bogi:18,zuo:19} for computing Tukey's depth in the Euclidean space.
This contrasts with the exact algorithms \citep[e.g.,][]{dyck:16,zuo:19} for computing Tukey's depth where the complexity is typically $O(n_{\cY}n^{m})$ or $O(n_{\cY}n^{m-1}\log(n))$. 
\ref{alg:deepest} takes $O(n^{3})$ since $n_\cY = n$.

\subsection{Theoretical Properties for the Approximation\label{ssec:approximationProperties}}

We establish that the approximate depth converges to the truth if the anchor points are dense enough in $\cM$. 
Halfspace $\Hzot$ is said to be a \emph{minimizing halfspace} at $x$ if $x \in \Hzot$ and $P_X(\Hzot)=D(x)$.
The following theorem derives the rate of convergence for the approximation if a minimizing halfspace exists, and the consistency result otherwise. 
{
To obtain the rate of convergence, for $z_j \in \cM$ let $D_{j}=d(X,z_{j})$, $j=1,2$ and assume the following conditions.

\begin{enumerate}[label=(P\arabic*),series=distributionD]
\item \label{a:probDj}For some $\epsilon > 0$ and $c_{1}>0$, $D_{j}$ has a small ball probability near 0 satisfying
$P(D_{j}\le t)\ge c_{1}t^{m_0}$ for $j=1,2$ and $t\le\epsilon$.
\item \label{a:condProbDj} For some $\epsilon > 0$ and $c_{2}>0$, $P(\left|D_{1}-D_{2}\right|\le t)\le c_{2}t$ holds for $t\le\epsilon$. 
\end{enumerate}

%\begin{enumerate}[label=(D\arabic*),series=distributionD]
%\item \label{a:density} Given $z_{1}\ne z_{2}\in\cM$, define $\hzot(t)=P(\left|d(X,z_{1})-d(X,z_{2})\right|\le t)$ 
%%as the distribution function of $\left|d(X,z_{1})-d(X,z_{2})\right|$
%and $\gzot(p)=\inf\{t\ge0\mid\hzot(t)\ge p\}$.
%% as the quantile function.
%We have  $\lim_{C\toinf}\lim_{\ntoinf}nP(d(X,z_{j})\le\gzot(Ca_{n})/2)=\infty$,
%for $j=1,2$ and some sequence $\{a_{n}\}$ with $\lim_{\ntoinf} a_n = 0$. 
%\end{enumerate}
\begin{thm}
\label{thm:tDnRate} Suppose that either \ref{a:uniformEntropy} or
\ref{a:bracketIntegral} holds, and the approximation algorithm uses
the sample points $\cX$ as the anchor points $\cA$. Let $x$ be
a point on $\cM$. 
\begin{enumerate}[label=(\alph*)]
\item \label{enu:tDnrootn}If the infimum in $D(x)=\inf_{H\in\cH_{x}}P_X(H)$
is achieved by a halfspace $\Hzot$, i.e., $D(x)=P_X(\Hzot)$, and \ref{a:probDj} and \ref{a:condProbDj}
hold for $(z_1, z_2)$, then as $\ntoinf$,
\[
\left|\tD_{n}(x)-D(x)\right|=O_{p}(n^{-1/m_0}).
\]
\item \label{enu:tDnop1}Suppose that the infimum of $D(x)=\inf_{H\in\cH_{x}}P_X(H)$
is not achieved by any halfspace. If $P(d(z,X)<r)>0$ for all $z\in\cM$
and $r>0$ and $P_X(\Eonetwo)=0$ for all $x_{1},x_{2}\in\cM$, then as $\ntoinf$,
\[
\left|\tD_{n}(x)-D(x)\right|=o_{p}(1).
\]
\end{enumerate}
\end{thm}
The idea of proof for \ref{thm:tDnRate} is to approximate the minimizing halfspace probabilities by random halfspaces. 
The halfspace where the infimum is attained does not need to be unique. 
Conditions \ref{a:probDj} and \ref{a:condProbDj} are requirements on both the distribution of $X$ and on the geometry of $\cM$.
They ensure that if the random anchor points lie close enough to the anchor points of a minimizing halfspace, then the halfspace probabilities are close. 
{} 
If $\cM$ is a Riemannian manifold and $X$ has a density bounded away from 0 w.r.t. the Riemannian volume measure, then $m_0$ in \ref{a:probDj} is the intrinsic dimension $m$ of $\cM$.
% \citep{lee:18} because the volume of a small geodesic ball
%scales as $r^{m}$ \citep[Theorem 3.1,][]{gray:74}. 
Thus, the rate of convergence of the approximation algorithm given by \ref{thm:tDnRate}\ref{enu:tDnrootn}  is as fast as $O_p(n^{-1/m})$ on an $m$-dimensional Riemannian manifold. 
%Condition \ref{a:probDj} and \ref{a:condProbDj} are requirements on both the distribution of $X$ and the geometry of $\cM$. 
Conditions \ref{a:probDj} and \ref{a:condProbDj} hold in a Euclidean space if $X$ has a regular density, 
and \ref{a:condProbDj} is violated if the distribution of $d(X,z_{1})$ is overly concentrated around $d(X,z_{2})$.
Two examples when \ref{a:probDj} and \ref{a:condProbDj} are satisfied and a counter-example are provided in \autoref{asec:conditionsExamples}, and additional properties of the approximate depth are described in \autoref{asec:propertiesApprox}.
}%

\section{Numerical Experiments\label{sec:numericalExperiments}}

%\subsection{Simulations on Riemannian Manifolds\label{ssec:simulationRiemannian}}

We investigate the performance of the metric halfspace median as a robust estimate for the center of a distribution. 
Three Riemannian manifolds were considered for the data space $\cM$, namely the $k\times k$ symmetric positive definite
matrices $\SPD(k)$ with the affine invariant metric; the $k$-dimensional
unit sphere $\bbS^{k}$; and the rotational group $\SO(k)$ of $k\times k$
orthogonal matrices with determinant 1%
\begin{comment}
; and the $k$-dimensional flat torus $\torusk=\bbS^{1}\times\dots\times\bbS^{1}$
\end{comment}
. The intrinsic dimensions $m$ for these manifolds equal, respectively,
$k(k+1)/2$, $k$, and $k(k-1)/2$.

For each metric space $\cM$, we considered four cases where i.i.d.
data were generated according to either an uncontaminated distribution
$\bbP=\bbP_{1}$ for Case 1 or contaminated distribution $\bbP=0.9\bbP_{1}+0.1\bbP_{2}$
for Cases 2 to 4. 
{
Population $\bbP_j$ followed the same distribution as $X_j=\exp_{\theta_j}(V_j)$, where $\exp_{\theta_j}$ is the Riemannian exponential map at a fixed center $\theta_j\in\cM$ and $V_j$ is a random tangent vector lying on $T_{\theta_{j}}\cM$, $j=1,2$;
additional details of the simulation setup are included in \ref{assec:setupsSimulations}.
Our target is to estimate robustly the center $\theta_1$ of the uncontaminated distribution $\bbP_{1}$, with the center and random tangent vector varying between simulation cases.
}%
% that is non-isotropic.
The contamination distribution $\bbP_{2}$ was set to
a distribution that differed from $\bbP_{1}$, in the location, scale,
and location-and-scale for Case 2, 3, and 4, respectively. To summarize,
the simulation scenarios considered were
\begin{itemize}
\singlespacing
\item Case 1, uncontaminated distribution centered at $\theta_1$,
\item Case 2, contaminated distribution with location outliers,
\item Case 3, contaminated distribution with scale outliers, and
\item Case 4, contaminated distribution with location-and-scale outliers.
\end{itemize}
{
For example, on $\bbS^2$, a location outlier is centered around $\theta_2$ that lies far away from the center $\theta_1$ of the uncontaminated distribution $\bbP_{1}$.
A scale outlier is generated from $V_2$ which has a different covariance matrix than $V_1$; therefore, a scale outlier  may lie away from its center in a direction uncommon to the inliers.
}%
%For example, a location outlier of rotational matrices on $\cM=\SO(k)$ differ from the inliers in the rotation
We varied the sample size $n\in\{50,100,200\}$ and the manifold parameter $k\in\{2,3,4\}$ in each case.

As estimators of the center, we compared the proposed metric halfspace
median $\hmu_{\MHD}=\rtheta$ as defined in (\ref{eq:ttheta}) and the Fréchet mean $\hmu_{\FM}$.
The Fréchet mean \citep{frec:48} of the sample $X_{i}$,
$i=1,\dots,n$ under distance $d$ is  $\hmu_{\FM}=\argmin_{x\in\cM}n^{-1}\sumin d^{2}(x,X_{i})$, which is a generalization of the classical mean. 
For calculating the metric halfspace median, 10 jiggled points were added to the anchor set around
each sample point. For $\cM=\SPD(k)$, we also compared with the Fréchet median $\hmu_{\GDD}=\argmin_{x\in\cM}n^{-1}\sumin d(x,X_{i})$, which is the deepest point w.r.t.\ the geodesic distance depth \citep{chau:19}. 
These location estimators $\hmu$ were evaluated according to the
median geodesic distance to the true mean $d(\hmu,\mu)$ out of {1024 Monte
Carlo repeats}.

Results for $\cM=\SPD(k)$ displayed in \ref{tab:affinv} show that
the proposed metric halfspace median performs well in general. In
Case 1 without contamination, the Fréchet mean was the most efficient overall, especially for smaller sample sizes $n=50$ and $100$, while
the metric halfspace median and the Fréchet median are competitive.
In the presence of contamination, both deepest points $\hmu_{\MHD}$
and $\hmu_{\GDD}$ dominated $\hmu_{\FM}$ and demonstrated robustness by producing estimates that were close in performance to
those in Case 1 without contamination. The proposed metric halfspace median
outperformed the Fréchet median in the contaminated scenarios. 
A reason for this is that the Fréchet median 
%sample geodesic distance depth $D_{n,\GDD}(x)=\exp\left(-n^{-1}\sumin d(x,X_{i})\right)$ \citep{chau:19} 
only considers the sum $\sumin d(x,X_{i})$ of geodesic distances from the data points to $x$, disregarding the relative locations of the data points within the point clouds and thus having weaker invariant properties than the metric
halfspace depth. 
The advantage of $\hmu_{\MHD}$ over $\hmu_{\GDD}$
becomes more significant when the sample size is larger, in which
case the approximation of the metric halfspace depth through $\tD_{n}$ is improved. 

\begin{table}[h]
\caption{Median distances over the replicates	between the estimated center and the actual center of $\bbP_{1}$ for data being symmetric positive definite matrices on $\protect\cM=\protect\SPD(k)$.
The dimensions of the manifold for parameters $k=2,3$, and $4$ are
$3,6,$ and $10$, respectively. 
{The standard errors of the reported median distances for $n=50,100,$ and $200$ were less than $0.003$, $0.002$, and $0.002$, respectively.}
MHD, the proposed metric halfspace
depth; FM, Fréchet mean; GDD, geodesic distance depth. \label{tab:affinv}}
\small
\centering
\begin{tabular}{lr@{\extracolsep{0pt}=}l|r@{\extracolsep{0pt}.}lr@{\extracolsep{0pt}.}lr@{\extracolsep{0pt}.}l|r@{\extracolsep{0pt}.}lr@{\extracolsep{0pt}.}lr@{\extracolsep{0pt}.}l|r@{\extracolsep{0pt}.}lr@{\extracolsep{0pt}.}lr@{\extracolsep{0pt}.}l}
 & \multicolumn{2}{c|}{} & \multicolumn{6}{c|}{{$k=2$}} & \multicolumn{6}{c|}{{$k=3$}} & \multicolumn{6}{c}{{$k=4$}}\tabularnewline
 & \multicolumn{2}{c|}{{$n=$}} & \multicolumn{2}{c}{{MHD}} & \multicolumn{2}{c}{{FM}} & \multicolumn{2}{c|}{{GDD}} & \multicolumn{2}{c}{{MHD}} & \multicolumn{2}{c}{{FM}} & \multicolumn{2}{c|}{{GDD}} & \multicolumn{2}{c}{{MHD}} & \multicolumn{2}{c}{{FM}} & \multicolumn{2}{c}{{GDD}}\tabularnewline
\hline 
\multirow{3}{*}{{Case 1}} & \multicolumn{2}{c|}{{50}} & &117   & & 103  & & 122   & & 116   & & 103  & & 121   & & 114   & & 103  & & 121 \tabularnewline
 & \multicolumn{2}{c|}{{100}} & & 075   & & 071  & & 081   & & 076   & & 071  & & 080   & & 078   & & 071  & & 080 \tabularnewline
 & \multicolumn{2}{c|}{{200}} & & 053   & & 049  & & 054   & & 053   & & 048  & & 054   & & 055   & & 048  & & 054 \tabularnewline
\hline 
\multirow{3}{*}{{Case 2}} & \multicolumn{2}{c|}{{50}} & & 124   & & 140  & & 136   & & 114   & & 140  & & 124   & & 119   & & 145  & & 126 \tabularnewline
 & \multicolumn{2}{c|}{{100}} & & 091   & & 120  & & 101   & & 084   & & 121  & & 097   & & 083   & & 121  & & 094 \tabularnewline
 & \multicolumn{2}{c|}{{200}} & & 070   & & 108  & & 084   & & 064   & & 110  & & 079   & & 059   & & 110  & & 077 \tabularnewline
\hline 
\multirow{3}{*}{{Case 3}} & \multicolumn{2}{c|}{{50}} & & 104   & & 107  & & 108   & & 102   & & 108  & & 104   & & 103   & & 105  & & 105 \tabularnewline
 & \multicolumn{2}{c|}{{100}} & & 072   & & 075  & & 074   & & 064   & & 075  & & 071   & & 063   & & 075  & & 071 \tabularnewline
 & \multicolumn{2}{c|}{{200}} & & 051   & & 054  & & 051   & & 042   & & 053  & & 050   & & 040   & & 054  & & 050 \tabularnewline
\hline 
\multirow{3}{*}{{Case 4}} & \multicolumn{2}{c|}{{50}} & & 123   & & 142  & & 133   & & 122   & & 145  & & 124   & & 120   & & 144  & & 125 \tabularnewline
 & \multicolumn{2}{c|}{{100}} & & 087   & & 124  & & 102   & & 084   & & 124  & & 094   & & 083   & & 125  & & 091 \tabularnewline
 & \multicolumn{2}{c|}{{200}} & & 065   & & 110  & & 086   & & 061   & & 112  & & 078   & & 062   & & 113  & & 077 \tabularnewline
\end{tabular}{ }
\end{table}

Results for two bounded manifolds are shown in \ref{tab:other}, where the exponential maps are not injective on these manifolds and thus depth concepts cannot be defined in general through mapping data onto the tangent space. 
The metric halfspace median is overall superior to the Fréchet mean
in the presence of contamination, especially when the intrinsic dimension
is large. Even in Case 3, where the scale-only outliers do not affect the true center, the metric halfspace median was, in many cases, more efficient than the Fréchet mean on spheres. This could be due to the low rate of convergence of the sample Fréchet mean for data that extends the entire manifold \citep{eltz:19}. 
Results for these different manifolds demonstrate that the proposed metric halfspace median is in general a valid robust measure of centrality.

%even in the presence of curvature and finite radius of injectivity. 
\begin{table}[h]
\caption{Median distances over the replicates	between the estimated and the actual centers
when data lie on the sphere $\protect\bbS^{k}$ and the rotational
group $\protect\SO(k)$. 
{The standard errors of the reported median distances for $n=50,100,$ and $200$ were less than $0.005$, $0.003$, and $0.004$, respectively.}
MHD, the proposed metric halfspace depth; FM, Fréchet mean.\label{tab:other}}
\begingroup\tabcolsep=3pt

\centering
\small
\begin{tabular}{lr@{\extracolsep{0pt}=}l|r@{\extracolsep{0pt}.}lr@{\extracolsep{0pt}.}l|r@{\extracolsep{0pt}.}lr@{\extracolsep{0pt}.}l|r@{\extracolsep{0pt}.}lr@{\extracolsep{0pt}.}l|r@{\extracolsep{0pt}.}lr@{\extracolsep{0pt}.}l|r@{\extracolsep{0pt}.}lr@{\extracolsep{0pt}.}l|r@{\extracolsep{0pt}.}lr@{\extracolsep{0pt}.}l}
 & \multicolumn{2}{c}{} & \multicolumn{12}{c|}{{$\cM=\bbS^{k}$}} & \multicolumn{12}{c}{{$\cM=\SO(k)$}}\tabularnewline
 & \multicolumn{2}{c|}{} & \multicolumn{4}{c|}{{$k=2$}} & \multicolumn{4}{c|}{{$k=3$}} & \multicolumn{4}{c|}{{$k=4$}} & \multicolumn{4}{c|}{{$k=2$}} & \multicolumn{4}{c|}{{$k=3$}} & \multicolumn{4}{c}{{$k=4$}}\tabularnewline
 & \multicolumn{2}{c|}{{$n=$}} & \multicolumn{2}{c}{{MHD}} & \multicolumn{2}{c|}{{FM}} & \multicolumn{2}{c}{{MHD}} & \multicolumn{2}{c|}{{FM}} & \multicolumn{2}{c}{{MHD}} & \multicolumn{2}{c|}{{FM}} & \multicolumn{2}{c}{{MHD}} & \multicolumn{2}{c|}{{FM}} & \multicolumn{2}{c}{{MHD}} & \multicolumn{2}{c|}{{FM}} & \multicolumn{2}{c}{{MHD}} & \multicolumn{2}{c}{{FM}}\tabularnewline
\hline 
\multirow{3}{*}{{Case 1}} & \multicolumn{2}{c|}{{50}}  &  & 146 & & 127 & & 144 & & 132 & & 147 & & 132 & & 120 & & 098 & & 128 & & 107 & & 133 & & 107 \tabularnewline
                          & \multicolumn{2}{c|}{{100}} &  & 096 & & 091 & & 096 & & 092 & & 098 & & 092 & & 082 & & 068 & & 091 & & 076 & & 095 & & 076 \tabularnewline
                          & \multicolumn{2}{c|}{{200}} &  & 070 & & 063 & & 069 & & 064 & & 070 & & 064 & & 057 & & 047 & & 083 & & 052 & & 081 & & 052 \tabularnewline
\hline
\multirow{3}{*}{{Case 2}} & \multicolumn{2}{c|}{{50}}  &  & 155 & & 152 & & 164 & & 170 & & 158 & & 172 & & 144 & & 125 & & 142 & & 149 & & 133 & & 147 \tabularnewline
                          & \multicolumn{2}{c|}{{100}} &  & 117 & & 133 & & 115 & & 137 & & 110 & & 141 & & 101 & & 105 & & 118 & & 124 & & 107 & & 125 \tabularnewline
                          & \multicolumn{2}{c|}{{200}} &  & 095 & & 113 & & 087 & & 120 & & 084 & & 124 & & 092 & & 098 & & 119 & & 110 & & 100 & & 112 \tabularnewline
\hline
\multirow{3}{*}{{Case 3}} & \multicolumn{2}{c|}{{50}}  &  & 137 & & 137 & & 140 & & 147 & & 130 & & 146 & & 120 & & 098 & & 119 & & 114 & & 111 & & 112 \tabularnewline
                          & \multicolumn{2}{c|}{{100}} &  & 097 & & 096 & & 091 & & 104 & & 089 & & 106 & & 082 & & 068 & & 087 & & 080 & & 076 & & 079 \tabularnewline
                          & \multicolumn{2}{c|}{{200}} &  & 069 & & 069 & & 066 & & 076 & & 059 & & 077 & & 057 & & 047 & & 075 & & 057 & & 066 & & 056 \tabularnewline
\hline
\multirow{3}{*}{{Case 4}} & \multicolumn{2}{c|}{{50}}  &  & 152 & & 167 & & 154 & & 182 & & 153 & & 187 & & 144 & & 125 & & 136 & & 149 & & 137 & & 149 \tabularnewline
                          & \multicolumn{2}{c|}{{100}} &  & 114 & & 146 & & 112 & & 157 & & 109 & & 157 & & 101 & & 105 & & 113 & & 129 & & 101 & & 125 \tabularnewline
                          & \multicolumn{2}{c|}{{200}} &  & 091 & & 128 & & 083 & & 135 & & 081 & & 135 & & 092 & & 098 & & 114 & & 113 & & 091 & & 112 \tabularnewline
\end{tabular}

\endgroup
\end{table}

\section{Real Data Applications\label{sec:applications}}

\subsection{Functional Connectivity in Alzheimer's Disease Patients\label{ssec:alzheimer}}

The first data application considers symmetric positive definite (SPD)
matrices that represent brain connectivity, which are widely used as a biomarker of brain function. The connectivity between defined regions of interest is calculated as the temporal association between
their blood-oxygen-level-dependent (BOLD) signals in functional magnetic resonance imaging (fMRI) scans when the subjects are in a resting state. 
%The connectivity of a subject is thus represented by a SPD covariance matrix. 
%A sample of connectivity matrices among dementia patients and healthy controls is then used to infer differences between pathological and normal brains.
We analyzed fMRI scans recorded in the Alzheimer's
Disease Neuroimaging Initiative (ADNI) with the goal of making inference
regarding brain connectivity in different dementia study groups.
Our analysis included $n=181$ subjects who, according to the severity of cognitive decline, were
classified at enrollment as: cognitively normal (CN), early mild cognitive impaired (EMCI), late mild cognitive impaired (LMCI), or Alzheimer's disease (AD) patients.
The fMRI data were preprocessed by following a standard protocol
to remove motion and timing artifacts, scaling effects, and trends,
and we considered only the fMRI scans at the participants' first visits.
Problematic scans are not uncommon in fMRI studies as a result of imaging artifacts that come from head motion and cognitive state \citep{laum:17}. 
Statistical depth approaches are appealing for analyzing imaging data since they are fully nonparametric and robust to outliers. 
Here we compare the proposed metric halfspace depth with
the geodesic distance depth \citep{chau:19}. 

\begin{figure}[h!]
\includegraphics[height=0.3\paperheight]{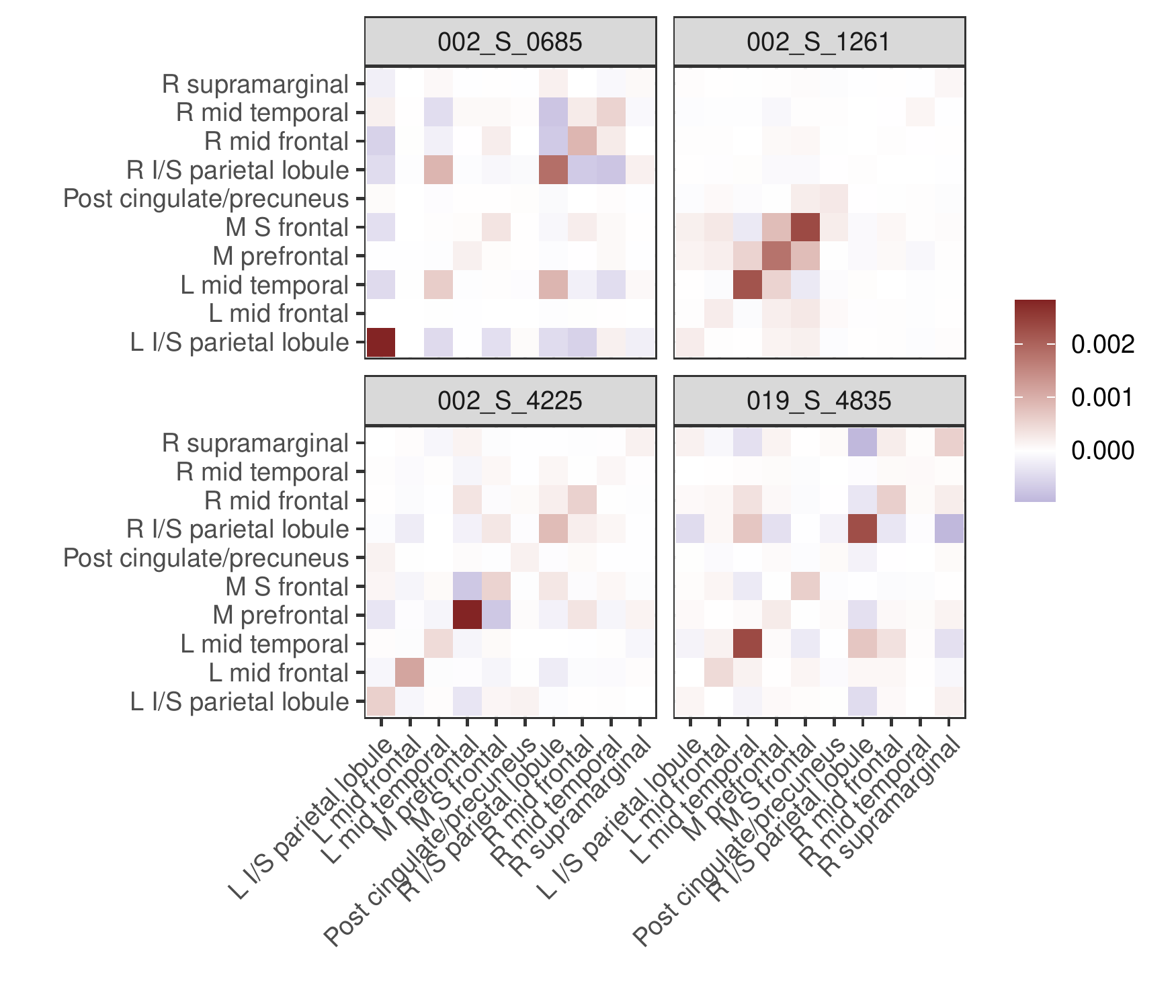}\includegraphics[height=0.3\paperheight]{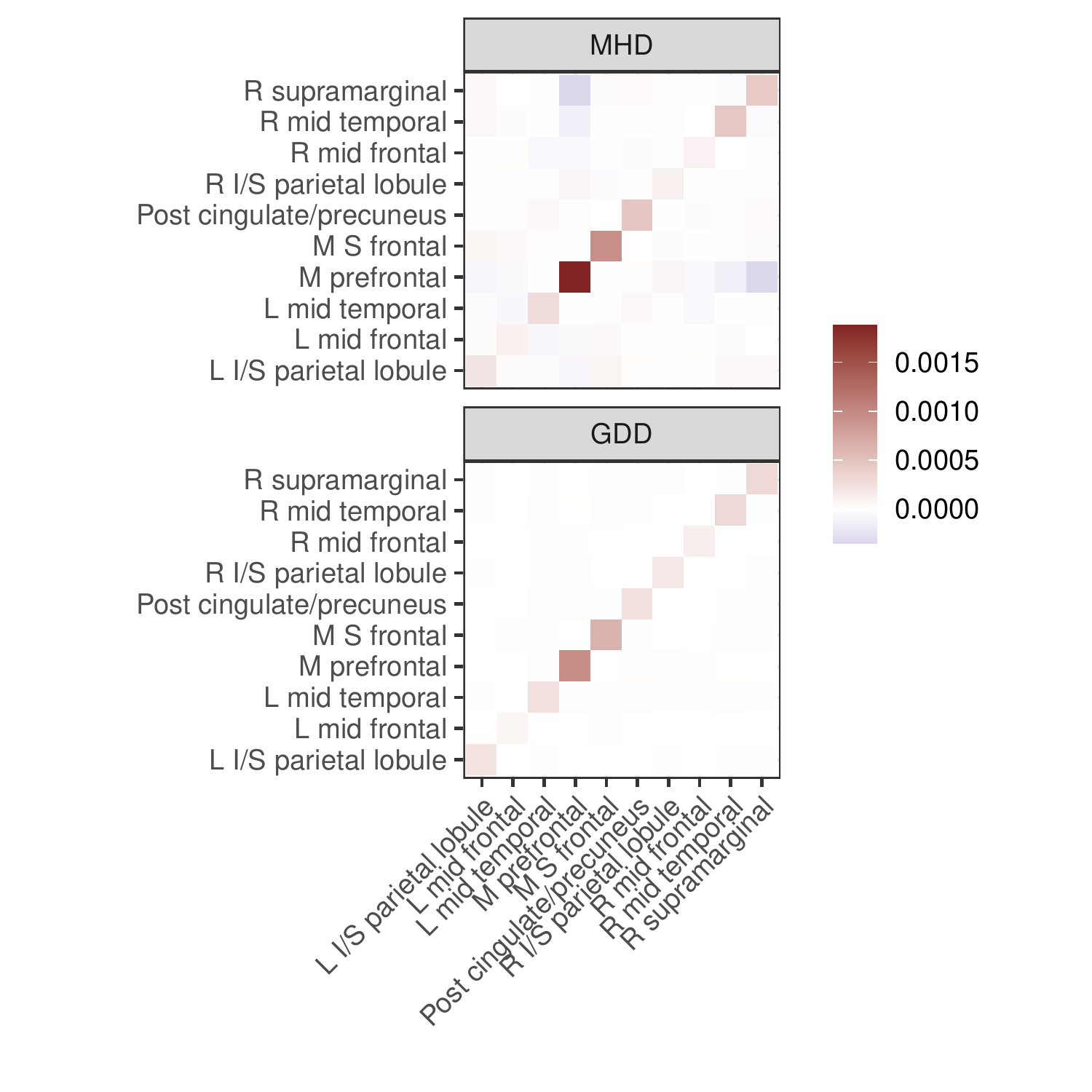}

\caption{Left: Connectivity covariance matrices for four cognitively normal individuals. Right:
Deepest matrices among cognitively normal individuals in terms of metric halfspace depth (MHD, upper panel) and
geodesic distance depth (GDD, lower panel). Brain regions used for creating the connectivity matrices are indicated.\label{fig:adni}
}
\end{figure}

For each subject, the average bold signals in each of the 10 defined brain regions (Buckner's hubs) in a subject's brain were first calculated, obtaining a 10-dimensional times series \citep{buck:09}. 
Next, brain connectivity is represented by the covariance (at lag 0) of the average bold signals, obtaining $10\times10$
covariance matrices as the data observations $X_{i}$. The left panel
of \ref{fig:adni} illustrates the connectivity covariance matrices
of four random subjects in the cognitively normal group. 
We analyzed the covariance matrices in $\cM=\SPD(10)$ with the affine invariant metric. 
The deepest covariance matrix in the cognitively normal group with respect to metric halfspace depth (upper right panel of \ref{fig:adni})
exhibits non-zero cross-covariances between different brain regions, resembling the original sample matrices; in contrast, the deepest image w.r.t. the geodesic distance depth \citep{chau:19} (lower right panel) has near 0 cross-covariances, which is not commonly observed in the sample. 

We next investigated whether group differences exist among the four groups of patients studied. 
We applied the depth-based Kruskal--Wallis test proposed by \cite{chen:12-2} based on both the proposed metric halfspace depth and the geodesic distance depth \citep{chau:19}. 
The Kruskal--Wallis test is designed to be sensitive to both location and scale changes by calculating the depth of the observations with respect to each group and aggregating the depth ranks. 
Using the permutation null distribution, a $p$-value of 0.0194 was produced using the proposed metric halfspace depth, and a $p$-value of 0.0652 for the geodesic distance depth. 
Further, pairwise comparisons of the dementia groups using the depth-based Wilcoxon test \citep{chen:12-2} 
revealed that the most significant difference exists between the Alzheimer's disease and the cognitively normal groups as shown in \ref{tab:ADNIpv}. 
This demonstrates the potential utility of fMRI-based connectivity measures and depth-based methods for studying Alzheimer's disease.

\begin{table}[h!]
\caption{The $p$-values based on the metric halfspace depth-based Wilcoxon rank test for the pairwise comparisons between the four dementia groups. 
CN, cognitively normal; EMCI, Early Mild Cognitive Impairment; LMCI, Late Mild Cognitive Impairment; AD, Alzheimer's disease.}
\small
\label{tab:ADNIpv}
\centering{}%
\begin{tabular}{c|ccc}
& EMCI & LMCI & AD \\ \hline
CN & 0.644 & 0.339 & 0.021 \\
EMCI & $-$ & 0.350 & 0.126 \\
LMCI & $-$ & $-$ & 0.074
\end{tabular}

\end{table}

\subsection{Phylogenetic Tree Application\label{ssec:api}}

\begin{comment}
Tree structured observations in the form of phylogenetic trees are
routinely investigated in bioinformatics, and these trees are data
objects lying on the tree space $\bbT^{k}$, a stratified geodesic
space that is not a Riemannian manifold. 
\end{comment}
%The second application considers phylogenetic trees. 
In evolutionary
biology, the ancestral relationship among a fixed collection of species
is represented by a tree structure. Each leaf corresponds to a species,
each interior node a speciation event, an edge the transition from
an ancestor to a descendant, and the edge length the evolutionary
divergence along the edge. A phylogenetic tree is constructed by comparing
genetic materials from different species and determining the divergence time from the mismatches between nucleic acid sequences. 
Frequently, a collection of phylogenetic trees are considered,
where each individual tree is constructed from the sequence of a specific gene present in the species in question. %\citep{kapl:20}. 
Collectively, this forms a sample of gene trees where the sources of randomness come from biological variation, sequence misalignment, and random subsampling in the individual genes. 
%Statistical methods
%for obtaining a consensus inference of the evolutionary history out
%of the sample of individual trees are thus very much needed \citep{nye:17,bard:18}. 

It has been of great interest to construct a consensus tree that summarizes
the individual trees to infer the evolutionary history. 
In addition to the complex structure of the trees, this task is complicated by
the stark heterogeneity in the individual trees due to analytic artifacts
such as sequence misalignment, remarkable biological variation, or
low signal-to-noise ratio in the random subsample. Recently, tree
space geometry-aware methods such as the Fréchet mean tree \citep[e.g.][]{nye:17}
have been proposed. 
These methods have been shown to produce reliable inference of tree topology and edge lengths. However, a preliminary outlier removal step \citep[e.g.,][]{weye:14} is usually performed since the Fréchet mean is a non-robust measure of location.
Here, we apply the metric halfspace depth to obtain a ``summary tree'' that best represents the data and to identify potential outliers. 
We infer the phylogeny of 7 pathogenic Apicomplexan species relative to an outgroup species using $n=268$ individual gene trees constructed by \citet{kuo:08}. 
The Apicomplexa phylum contains many important
pathogenic parasites that are detrimental to humans and livestock. 
The Apicomplexan species included the infamous\emph{ }malaria
pathogens \emph{Plasmodium falciparum} (Pf) and \emph{Plasmodium vivax} (Pv);
tick-borne haemopathogens \emph{Babesia bovis} (Bb) and \emph{Theileria
annulata} (Ta); and coccidian parasites \emph{Eimeria tenella} (Et), \emph{Toxoplasma gondii} (Tg), and \emph{Cryptosporidium parvum} (Cp) which infect intestines. 
The outgroup \emph{Tetrahymena thermophila} (Tt) is a remotely related model species 
%in the phylum Ciliophora 
included to root the phylogeny. 
We model the gene
trees as rooted trees with the root placed as the point where the
outgroup joins with the apicomplexan species. 

%The BHV-tree space considers
%the lengths of only the interior edges since the $k$ pendant edges
%connecting the leaves in a $k$-tree have relatively simple structure
%and can be represented by a point in $\bbRnn^{k}$. 

\begin{figure}[h]
\centering{}\includegraphics[width=0.5\textwidth]{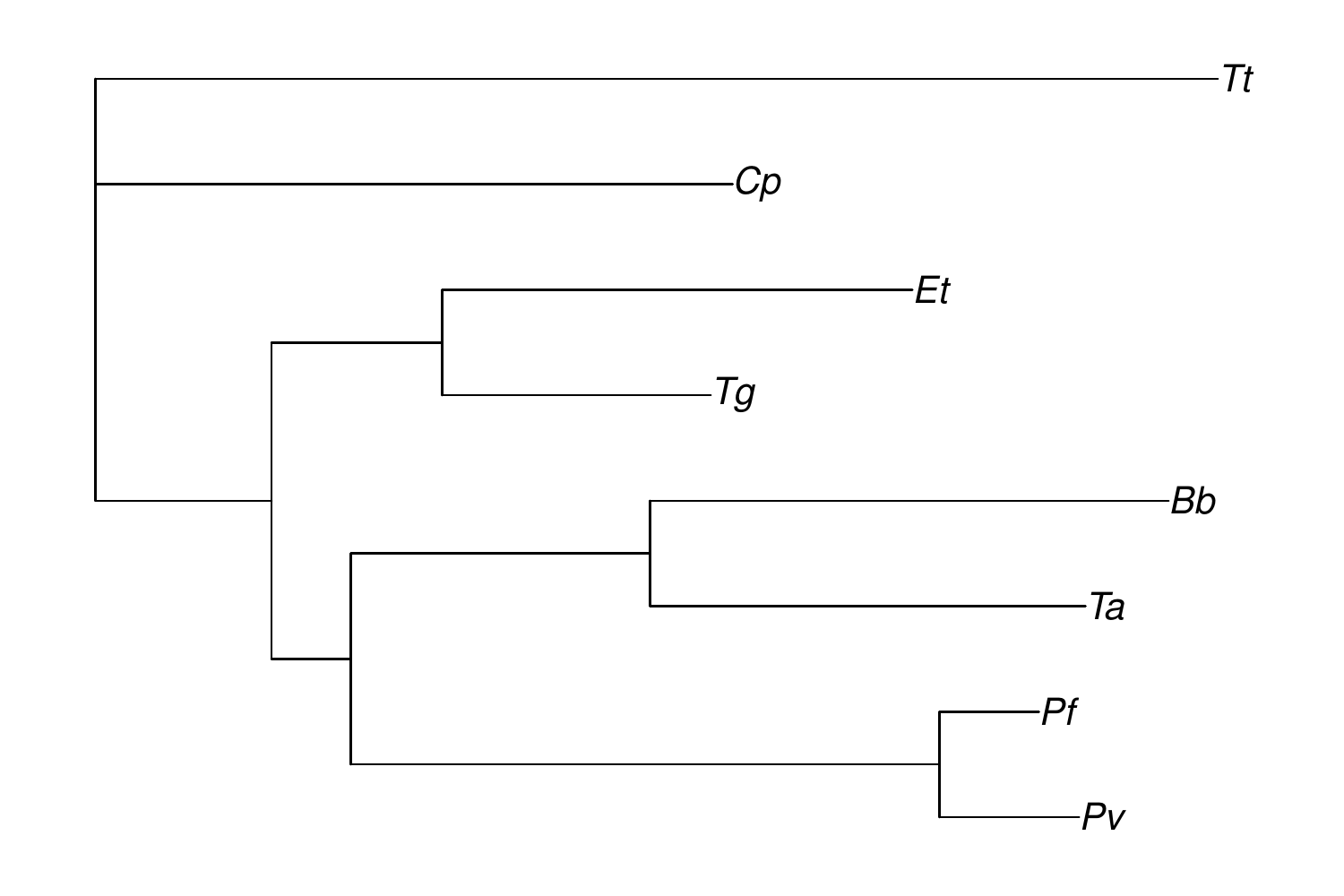}\caption{The deepest tree with respect to the proposed metric halfspace depth. 
The tree topology coincides with the known topology for the apicomplexan species tree. 
\label{fig:deepestTree}}
\end{figure}

To model the evolutionary divergence between all species and their ancestors, we consider $\bbT^{8}\times\bbR^{8}$ with the product metric, where the BHV space $\bbT^8$ models the tree topology and the interior edge lengths, and $\bbR^8$ models the pendant edge lengths. 
The proposed metric halfspace depths were calculated at each of the
individual trees, with 10 additional jiggled trees added as anchors per original tree for improving approximation. 
In the deepest tree 
as displayed in \ref{fig:deepestTree}, tick parasites \emph{B. bovis}
and \emph{T. annulata} and malaria parasites \emph{P.} \emph{falciparum}
and \emph{P. vivax} are respectively monophyletic, i.e., sharing the same immediate ancestor; 
these haemoparasites descend from a common ancestor; 
coccidian species \emph{E. tenella} and \emph{T. gondii} form a sister group to the former; 
\emph{C. parvum} is the deepest rooting species. 
The deepest tree we produced is congruent to the consensus tree identified by \citet{kuo:08} constructed through maximum likelihood, maximum
parsimony, and neighbor-joining methods, and also agree with the Fréchet
mean tree found by \citet{nye:17}, who performed the analysis after removing 16 outliers. 
Our depth-based approach has the advantage of being robust to extreme values and does not require separate outlier identification and removal.

\begin{figure}[h]
\centering{}\includegraphics[width=1\textwidth]{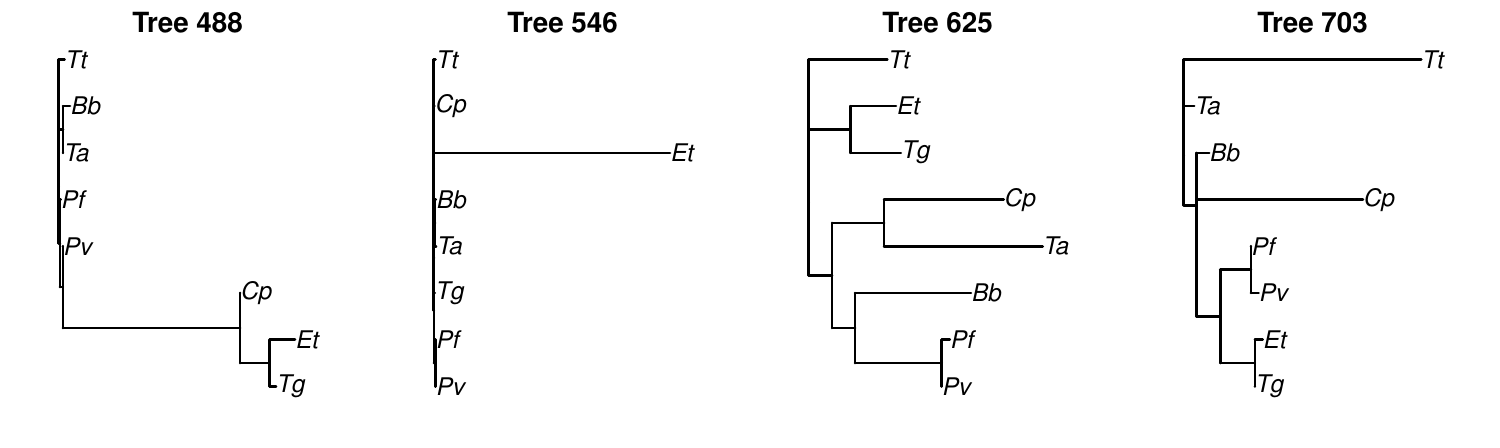}
\caption{Four individual gene trees with the least metric halfspace depth. \label{fig:outlierTrees}}
\end{figure}

We also identified 27 gene trees with the least metric halfspace depth, indicating that they correspond to the most extreme trees. 
Among these trees four potential outliers are displayed in \ref{fig:outlierTrees} and
the rest are included in \ref{fig:allOutlierTrees}. 
Trees 488 and 546 
have exceptionally long branches, and, in addition, the \emph{Plasmodium} species in tree 488 (Pf and Pv, hard to distinguish in the figure due to the long branch) and tick parasites \emph{B. bovis} and \emph{T. annulata} in trees 625 and 703  are not monophyletic. These structures, which differ from what has been reported in the literature \citep{kuo:08}, demonstrate the utility of the metric halfspace depth for highlighting outliers. 

\references

\appendix

\section{Riemannian Manifolds \label{asec:riemannianManifold}}

A subset $\cM\subset\bbR^{m+k}$ is an $m$-dimensional \emph{submanifold
}if, for any $x\in\cM$, there exists a neighborhood $U$ of $x$
in $\bbR^{m+k}$ and a smooth function $f:U\rightarrow\bbR^{k}$ with
everywhere surjective differential map such that $U\cap\cM=f^{-1}(0)$.
We say that $m$ is the \emph{intrinsic dimension} and $m+k$ the \emph{ambient
dimension}. 
The \emph{tangent space $T_{x}\cM\subset\bbR^{m+k}$} at $x\in\cM$ is an $m$-dimensional subspace consists of \emph{tangent vectors} of the form $\alpha'(0)$ where $\alpha:(-\epsilon,\epsilon)\rightarrow\cM$ is a differentiable curve on $\cM$ passing through $\alpha(0)=x$ for some $\epsilon>0$. 
A \emph{Riemannian manifold} is a submanifold
of $\bbR^{m+k}$ endowed with the \emph{Riemannian metric} $\inner{\cdot}{\cdot}_{x}:T_{x}\cM\times T_{x}\cM\rightarrow\bbR$
at each $x\in\cM$ that equals to the inner product in the ambient
Euclidean space. The \emph{Riemannian length} $l(\alpha)$ of a piecewise
differentiable path $\alpha:[a,b]\rightarrow\cM$ is defined by $l(\alpha)=\int_{a}^{b}\inner{\alpha'(t)}{\alpha'(t)}_{x}^{1/2}dt$.
The \emph{Riemannian distance} between $x,y\in\cM$ is the infimum
over the length $l(\alpha)$ of piecewise differentiable paths $\alpha:[0,1]\rightarrow\cM$
with $\alpha(0)=x$ and $\alpha(1)=y$. If $\cM$ is a complete connected
Riemannian manifold, by the Hopf--Rinow theorem \citep{lee:18}, there
exists a geodesic between any two points $x,y\in\cM$, so $\cM$ is
a geodesic space as well. 

For generating data on Riemannian manifolds, let $\exp_{x}:T_{x}\cM\rightarrow\cM$
be the \emph{Riemannian exponential map} at $x\in\cM$, where $\exp_{x}(v)$
is defined as the endpoint $\gamma(1)$ of a unit-speed geodesic $\gamma:[0,1]\rightarrow\cM$
with $\gamma(0)=x$, $\gamma'(0)=v$, and $\norm{\gamma'(t)}=1$ for
all $t$. 
The exponential map takes different forms on different Riemannian manifolds; some examples can be found in the Supplemental Materials.
{A well-defined \emph{inverse exponential map} $\exp_x^{-1}$ exists as a function mapping from $\cM$ to $T_x\cM$ if $\exp_x$ is injective. }

% Acknowledgement: Reviewers, Kuo for sharing data, grants
\end{document}

%% file: preamble.tex
\PassOptionsToPackage{linesnumbered,lined,ruled}{algorithm2e}
\usepackage[utf8]{inputenc}
\usepackage[T1]{fontenc}
\usepackage{babel}
\usepackage{array}
\usepackage{verbatim}
\usepackage{float}
\usepackage{mathtools}
\usepackage{enumitem}
\usepackage{multirow}
\usepackage{algorithm2e}
\usepackage{amsmath}
\usepackage{amsthm}
\usepackage{amssymb}
\usepackage{graphicx}
\usepackage{rotfloat}
\usepackage{setspace}
\usepackage[authoryear]{natbib}
\usepackage{xr-hyper}
\usepackage[unicode=true,pdfusetitle,
 bookmarks=true,bookmarksnumbered=false,bookmarksopen=false,
 breaklinks=false,pdfborder={0 0 1},backref=false,colorlinks=false]
 {hyperref}
\hypersetup{
colorlinks,allcolors=blue}
\makeatletter
\newcommand*{\addFileDependency}[1]{% argument=file name and extension
  \typeout{(#1)}
  \@addtofilelist{#1}
  \IfFileExists{#1}{}{\typeout{No file #1.}}
}
\newcommand*{\myexternaldocument}[1]{%
    \externaldocument{#1}%
    \addFileDependency{#1.tex}%
    \addFileDependency{#1.aux}%
}
\makeatother

\makeatletter

\providecommand{\tabularnewline}{\\}
\theoremstyle{definition}
 \newtheorem{example}{\protect\examplename}
\theoremstyle{plain}
\newtheorem{thm}{\protect\theoremname}
\theoremstyle{plain}
\newtheorem{prop}{\protect\propositionname}
\theoremstyle{plain}
\newtheorem{cor}{\protect\corollaryname}
\theoremstyle{plain}

\@ifundefined{date}{}{\date{}}
\usepackage{enumitem}
\usepackage{ifthen}
\usepackage{xcolor}

\usepackage{natbib} 
\usepackage{titlesec}
\titleformat{\section}[block]{\Large\bfseries}{\thesection}{1em}{}
\titleformat{\subsection}[block]{\large\bfseries}{\thesubsection}{1em}{}
\titlespacing*{\section}{0pt}{*1}{*1}
\titlespacing*{\subsection}{0pt}{*0}{*1}
\titlespacing*{\subsubsection}{0pt}{*1}{*1}

\def\LyX{\texorpdfstring{%
  L\kern-.1667em\lower.25em\hbox{Y}\kern-.125emX\@}
  {LyX}}

\newcommand{\mylabel}[2]{#2\def\@currentlabel{#2}\label{#1}}
\newcommand{\vertiii}[1]{{\left\vert\kern-0.25ex\left\vert\kern-0.25ex\left\vert #1 
    \right\vert\kern-0.25ex\right\vert\kern-0.25ex\right\vert}}

\AtBeginDocument{%
\let\ref\autoref
\renewcommand\equationautorefname{\@gobble}

}

\makeatother

\providecommand{\corollaryname}{Corollary}
\providecommand{\examplename}{Example}
\providecommand{\lemmaname}{Lemma}
\providecommand{\propositionname}{Proposition}
\providecommand{\theoremname}{Theorem}

\global\long\def\quadas{\quad a.s.}%
\global\long\def\bbH{\mathbb{\mathbb{H}}}%
\global\long\def\bbS{\mathbb{\mathbb{S}}}%
\global\long\def\bbT{\mathbb{T}}%
\global\long\def\bbR{\mathbb{R}}%
\global\long\def\bbP{\mathbb{P}}%
\global\long\def\bbH{\mathbb{H}}%

\global\long\def\transpose{\intercal}%

\global\long\def\hmu{\hat{\mu}}%
\global\long\def\htheta{\hat{\theta}}%
\global\long\def\ttheta{\tilde{\theta}}%
\global\long\def\rtheta{\mathring{\theta}}%
\global\long\def\tD{\tilde{D}}%

\global\long\def\cA{\mathcal{A}}%
\global\long\def\cB{\mathcal{B}}%
\global\long\def\cC{\mathcal{C}}%
\global\long\def\cF{\mathcal{F}}%
\global\long\def\cH{\mathcal{H}}%
\global\long\def\cM{\mathcal{M}}%
\global\long\def\cN{\mathcal{N}}%
\global\long\def\cS{\mathcal{S}}%
\global\long\def\cX{\mathcal{X}}%
\global\long\def\cY{\mathcal{Y}}%
\global\long\def\cZ{\mathcal{Z}}%

\global\long\def\logm{\operatorname{logm}}%

\global\long\def\argmax{\operatornamewithlimits{arg\,max}}%
\global\long\def\argmin{\operatornamewithlimits{arg\,min}}%

\global\long\def\toinf{\rightarrow\infty}%
\global\long\def\tozero{\rightarrow0}%
\global\long\def\ntoinf{n \rightarrow\infty}%
\global\long\def\one{I}%

\global\long\def\sumin{\sum_{i=1}^{n}}%

\global\long\def\norm#1{\left\lVert #1\right\rVert }%
\global\long\def\normF#1{\left\lVert #1\right\rVert _{F}}%
\global\long\def\inner#1#2{\langle#1,#2\rangle}%

\global\long\def\Nbrac{N_{[]}}%

\global\long\def\VC{\operatorname{VC}}%
\global\long\def\Tukey{\textrm{Tukey}}%

\global\long\def\Honetwo{H_{x_{1}x_{2}}}%
\global\long\def\Hyote{H_{y_{1}y_{2},e}}%
\global\long\def\Eonetwo{E_{x_{1}x_{2}}}%
\global\long\def\Hzot{H_{z_{1}z_{2}}}%
\global\long\def\supxM{\sup_{x\in\cM}}%

\global\long\def\comp{\mathrm{c}}%
\global\long\def\infxoxt{\inf_{\substack{x_{1},x_{2}\in\cM\\
 d(x_{1},x)\le d(x_{2},x) 
}
 }}%

\global\long\def\bbRnn{\bbR_{\ge0}}%
\global\long\def\torusk{(\bbS^{1})^{\times k}}%
\global\long\def\SPD{\operatorname{SPD}}%
\global\long\def\SO{\operatorname{SO}}%
\global\long\def\FM{\operatorname{FM}}%
\global\long\def\MHD{\operatorname{MHD}}%
\global\long\def\GDD{\operatorname{GDD}}%

\global\long\def\cXn{\cX^{(n)}}%
\global\long\def\cYl{\cY^{(l)}}%